\documentclass[12pt]{article}
\newif\ifams\amsfalse                                                    
\amstrue                                                                 
\ifams                                                                   
 \usepackage{amsfonts}                                                   
\fi                                                                      
\newif\iffigs\figsfalse                                                  
\figstrue                                                                
\newif\ifdraft\draftfalse
\newif\ifinter\interfalse
\intertrue
\ifdraft\else\interfalse\fi
\ifinter
 \else
\fi
\title{}

\ifdraft
  \def\secl#1{\nopagebreak\marginpar{\vspace{-6mm}\scriptsize #1}\label{#1}}
  \def\beql#1{\marginpar{\vspace{4mm}\scriptsize #1}
              \nopagebreak\begin{equation}\label{#1}}
  \def\ftl#1#2{\footnote{\label{#1}[#1] #2}}
  \def\capl#1#2{\caption{[#1] #2}\label{#1}}
  \def\bibl#1{\marginpar{\vspace{4mm}\scriptsize #1}\nopagebreak\bibitem{#1}}
 \else
  \def\secl#1{\label{#1}}
  \def\beql#1{\begin{equation}\label{#1}}
  \def\ftl#1#2{\footnote{\label{#1}#2}}
  \def\capl#1#2{\caption{#2}\label{#1}}
  \def\bibl#1{\bibitem{#1}}
\fi

\def\cit#1{\ifdraft[#1]\fi}

\def\draftnote#1%
  {\ifdraft
    \hspace{1mm}\newline\noindent\begin{tabular}[t]{|p{14cm}|}
     \hline \rule{0mm}{2.5ex} \underline{DRAFT NOTE}: #1 \\ \hline
    \end{tabular}\newline\noindent
   \fi}

\def\internote#1%
  {\ifinter
    \hspace{1mm}\newline\noindent\begin{tabular}[t]{|p{14cm}|}
     \hline \rule{0mm}{2.5ex} \underline{Internal Note}: #1 \\ \hline
    \end{tabular}\newline\noindent
   \fi}

\def\multdn{$\downarrow\downarrow\downarrow\downarrow\downarrow$}
\def\beginsup%
 {\noindent\begin{tabular}[t]{|c|}
   \hline  Supplements\\
   \multdn\multdn\multdn\multdn\multdn\multdn
   \multdn\multdn\multdn\multdn\multdn\multdn \\ \hline
  \end{tabular}}

\def\multup{$\uparrow\uparrow\uparrow\uparrow\uparrow$}
\def\endsup%
 {\noindent\begin{tabular}[t]{|c|}
   \hline \multup\multup\multup\multup\multup\multup
          \multup\multup\multup\multup\multup\multup \\
   Supplements\\ \hline
  \end{tabular}}

\iffigs
 \input epsf
 \def\pct#1{\centerline{ \epsfbox{#1.eps}}}
\else
 \def\pct#1{(see figure in file #1.eps)}
\fi
\newif\ifappend\appendfalse
\appendtrue
\ifappend
 \newcommand{\newsection}[1]{
  \vspace{7mm} \pagebreak[3]
  \refstepcounter{section}
  \setcounter{equation}{0}
  \message{(\thesection. #1)}
  \addcontentsline{toc}{section}{
   \protect\numberline{\thesection}{\hs\hs\boldmath #1}}
  \begin{flushleft}
   {\large\bf \thesection. #1}
  \end{flushleft}
  \nopagebreak}

\else
 \newcommand{\newsection}[1]{\section{#1}}
\fi

\newcommand{\newpar}[1]{
 \vspace{3mm}
 \noindent{\bf #1}
 \vspace{2mm} \nopagebreak}

\def\al{\alpha}
\def\bt{\beta}
\def\gm{\gamma}                \def\Gm{\Gamma}
\def\dl{\delta}                \def\Dl{\Delta}
\def\ep{\epsilon}

\def\th{\theta}               
\def\vph{\varphi}

               \def\Om{\Omega}
\def\sg{\sigma}               

\def\Ec{\mbox{\protect$\cal E$}}
\def\Oc{\mbox{\protect$\cal O$}}

\ifams
 \def\bbl#1{{\mathbb #1}}
\else
 \def\bbl#1l{{\bf #1}}
\fi
\def\CC{\bbl{C}}

\def\RR{\bbl{R}}
\def\ZZ{\bbl{Z}}

\def\max{{\rm max}}

\def\mod{{\rm \hspace{0.5ex} mod \hspace{0.5ex}}}

\def\eff{{\rm eff}}

\def\pt{\partial}
\def\goto{\rightarrow}
\def\Goto{\Rightarrow}

\def\wbar{\overline}

\def\inv{^{-1}}
\def\siml{\raisebox{-1ex}{$\stackrel{\textstyle <}{\sim}$}}

\def\rec#1{{\raise 0.4ex \hbox{$\scriptstyle {\frac{1}{#1}}$}}}

\def\half{{\raise 0.4ex \hbox{$\scriptstyle {1 \over 2}$}}}

\def\hs{\hspace{2mm}}

\def\hsc{\hspace{2mm},\hspace{5mm}}
\def\nl{\protect\newline}
\def\nlb{\protect\newline $\bullet$ }

\def\ie{{\em i.e.}}
\def\eg{{\em e.g.}}

\def\beq{\begin{equation}}
\def\eeq{\end{equation}}

\def\YM{{\rm YM}}

\def\prl{\parallel}

\begin{document}


\begin{titlepage}

\begin{flushright}
EFI-99-3\\ KUL-TF-99/06\\ hep-th/9902045\\
\end{flushright}

\ifinter \vspace{-10mm} \else \vspace{5mm} \fi

\begin{center}
\LARGE {\bf The D2-D6 System \\ and \\ a Fibered AdS Geometry} \\

\ifinter \vspace{5mm} \else \vspace{10mm} \fi

\large Oskar Pelc$^{1}$ \normalsize and \large Ruud
Siebelink$^{1,2,\dagger}$ \normalsize \\

\vspace{5mm}

$^1$ {\em Enrico Fermi Institute\\ University of Chicago \\ 5640
S. Ellis Ave. Chicago, IL 60037, USA} \\

\vspace{5mm}

$^2$ {\em Instituut voor theoretische fysica, \\ Katholieke
Universiteit Leuven\\ B-3001 Leuven, Belgium} \\

\ifinter \else \vspace{5mm} \fi

E-mail: {\tt oskar,siebelin@theory.uchicago.edu}
\end{center}

\ifinter \else \vspace{10mm} \fi

\begin{center}\bf Abstract\end{center}
\begin{quote}
The system of D2 branes localized on or near D6 branes is considered.
The world-volume theory on the D2 branes is investigated, using its
conjectured relation to the near-horizon geometry.
The results are in agreement with known facts and expectations for the
corresponding field theory and a rich phase structure is obtained as a
function of the energy scale and the number of branes.
In particular, for an intermediate range of the number of D6 branes, the
IR geometry is that of an $AdS_4$ space fibered over a compact space.
This D2-D6 system is compared to other systems, related to it by
compactification and duality and it is shown that the
qualitative differences have compatible explanations in the geometric and
field-theoretic descriptions. Another system -- that of NS5 branes located at
D6 branes -- is also briefly studied, leading to a similar phase structure.
\end{quote}


\ifinter \else \vspace{5mm} \fi
\hrule width 5.cm \vskip 2.mm {\small \noindent
$^\dagger$ Post-doctoraal Onderzoeker FWO, Belgium}

\internote{
\nl Keywords: SUSY GT, String Theory, D branes, AdS, SUGRA
\nl PACS codes: 11.10.Kk (FT in $D\neq4$), 11.15.-q (GT),
11.25.-w (Fund. Strings), 11.30Pb (SUSY).}
%

\end{titlepage}


\flushbottom


\newsection{Introduction and Summary}

In recent years, and especially since the discovery of the
so-called AdS-CFT correspondence
\cite{Mald9711,GKP9802,witten9802}, it has become clear that many
aspects of field theories can be studied by looking at the
near-horizon region of brane geometries.
In his pioneering work \cite{Mald9711}, Maldacena considered D3,
M2 and M5 branes, as well as the D1 - D5 intersection and
suggested a novel duality: on the one hand there is the field theory
describing the low-energy dynamics of the brane configuration,
when it decouples from the bulk stringy degrees of freedom;  and it is 
conjectured to be dual to
string/M theory in the geometry near the
horizon of the branes.
This is a strong-weak duality: whenever one of the dual
descriptions is weakly coupled and can be treated
semi-classically, the other is strongly coupled and the quantum
corrections are large. Specifically, when the number of branes is
large (and, in string theory, also when the string coupling is
small) the string/M theory in the above cases is well approximated
by classical supergravity and this can be used to learn about the
dual, strongly-coupled, field theory.

For each of the systems mentioned above, the brane configuration has a
near-horizon geometry of the $AdS_{n+1} \times S^m$ form, while
the corresponding $n$ dimensional dual field theory is conformally
invariant (hence the term AdS-CFT). One piece of evidence for
the duality is the match between the isometries of the
near-horizon geometry and the symmetries of the field theory. In
particular, the isometries of the $AdS_{n+1}$ factor are
identified with the conformal transformations in the field theory.
The AdS-CFT correspondence continues to hold when the $S^m$ factor
in the near-horizon geometry is replaced by another Einstein space
$X_m$. The killing spinors on $X_m$ determine the amount of
supersymmetry that is realized in the corresponding CFT and
spheres lead to the maximally supersymmetric cases. A large class
of examples with reduced supersymmetry can be obtained by
orbifolding those spheres \cite{orb}-\cite{GuRaWi}. A general
classification of possible Einstein spaces $X_m$, as well as the
resulting supersymmetries, has been given in
\cite{class,MoPl9810}. However, as we will see below, this product
structure is not the most general geometry that can correspond to
a conformal field theory.

The original Maldacena conjecture has been generalized also to
non-conformal theories. This
was first done in \cite{IMSY9802}, where it was shown how an
intricate phase structure emerges for the various branes in type
II string theory.
There, the distance from the brane was identified with the energy scale in
the world-volume field theory and it was found that
the 10 dimensional classical geometry associated with the branes can only
be trusted for a limited range of gauge theory 
energies.\footnote{excluding the D3 brane, which corresponds to a
{\em conformal} field theory.}
When going outside this range of validity two things can happen.
On the one hand the dilaton may diverge, which signals a transition towards
M-theory (for the type IIA branes) or to an S-dual regime (for the
type IIB cases); on the other hand, the curvature may become large, in which
case the geometric approach breaks down altogether. Fortunately,
this precisely happens when the gauge theory becomes weakly
coupled. In summary, one finds that for each energy scale there is
precisely one of the available dual descriptions that can be treated
semi-classically.
Nowhere do inconsistencies arise, e.g. due to two dual
descriptions being weakly coupled at the same energy scale.
The resulting phase structure for D2 branes is described in the left part of
figure \ref{f-phs}.

Although the evidence supporting Maldacena's duality is, by now,
quite impressive, there is still much to be understood. In order to
extend our understanding, it would be useful to investigate more
systems in which this duality exist and can be effectively
checked. In this work we follow this route. We concentrate
primarily on the D2-D6 system: a system of D2 branes located on or
near D6 branes. One of our main results is the phase structure of
the corresponding world-volume theory. It is described in figure
\ref{f-phs} and we explain it below.
\begin{figure}
\pct{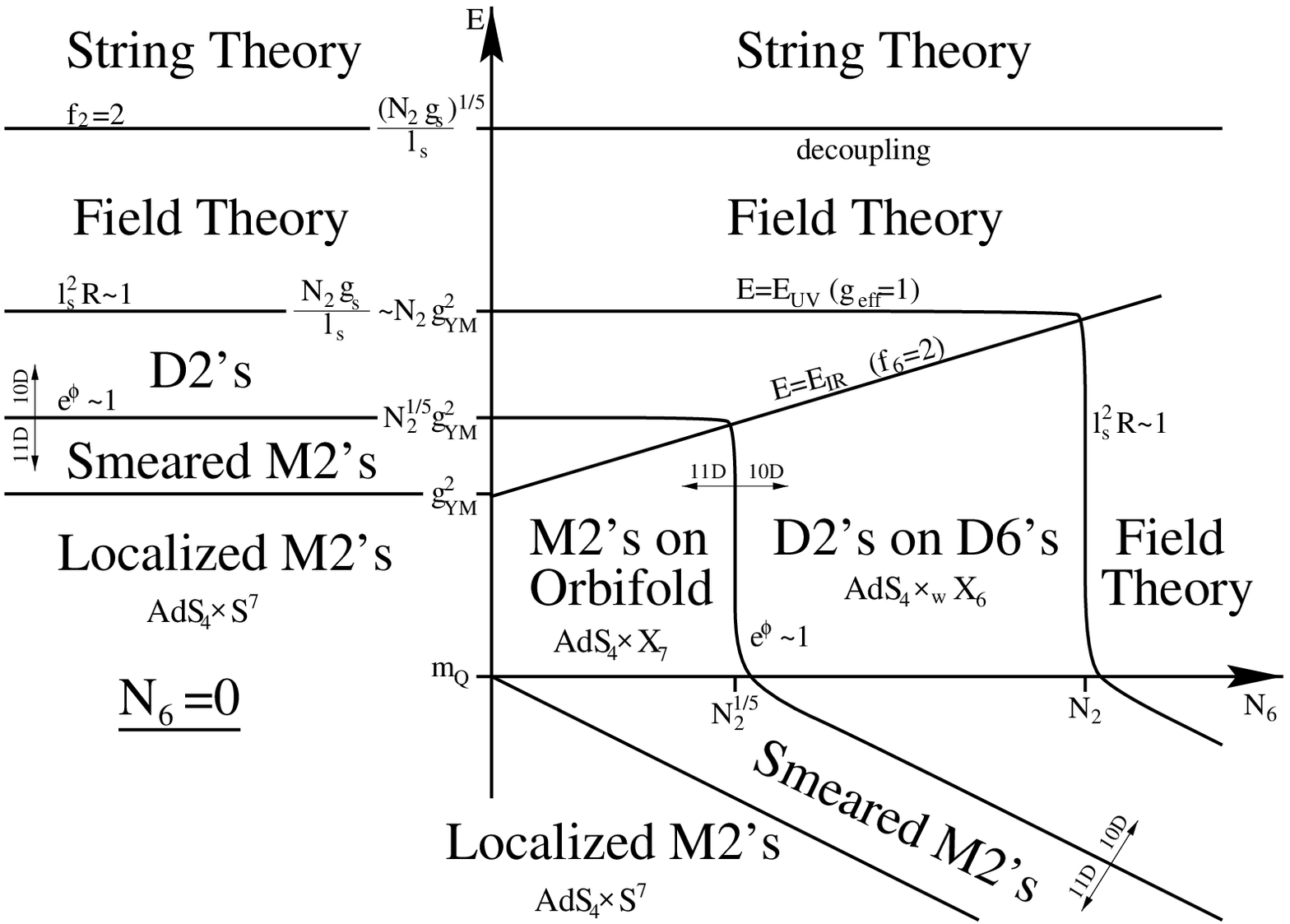}
\capl{f-phs}{The phase diagram of the world-volume theory}
\end{figure}

At low energies -- i.e. when the D2 brane dynamics decouples from the bulk --
the world-volume theory on the branes
is described by 3 dimensional $N=4$
supersymmetric gauge theory, with gauge group $SU(N_2)$ and $N_6$
hypermultiplets in the fundamental representation (``quarks''),
where $N_2$ and $N_6$ are the number of D2 and D6 branes, respectively.
Following Maldacena's conjecture, this theory is expected to be
dual to string theory in the geometry near the horizon of the D2 branes.
We argue that the relevant geometry is that of D2 branes
{\em localized} on the D6 branes.
This geometry is known only close enough to the D6 branes \cite{ITY3103}, 
or very far from them (where their influence is negligible).
We find that the transition between these two regions can occur     
within the near-horizon range of the D2 branes and, therefore, it is a
transition in the world-volume theory.
Indeed, the region close to the D6 branes corresponds to the range of
energy scales governed by the IR fixed point, as can be seen by noting that 
the near-horizon geometry we obtain is $SO(3,2)$ invariant.
We analyze quantitatively the geometric identification of the
field-theoretic energy scale, and determine the bound $E_{IR}$
on the range of the IR fixed point. 

Far enough from the D6 branes, where their influence on the
geometry can be neglected, one obtains the geometry of isolated D2 branes,
which was analyzed in \cite{IMSY9802}.
In particular, at a large enough distance from the D2 branes, the classical
geometric description stops being valid, due to a large curvature, while the
field-theoretical description becomes weakly coupled.
Comparing the two transition points $E_{IR}$ and $E_{UV}$,
one finds, as required, that they obey $E_{IR}\le E_{UV}$
whenever the geometric description is valid.
In fact, this is a strict inequality,
meaning that there is an intermediate phase.
One should note that the relevance of the geometry without the D6 branes
does not mean that in the world-volume theory, the quarks decouple at high
energies. The energy is related to the distance from the D2 branes and at
any such distance there are regions close to the D6 brane.
\internote{Comparing the geometric regions corresponding to high energies,
without D6 branes it is a spherical shell around the D2 branes, while with
D6 branes the regions close to them are ``cut out'' from the above sphere.}

Restricting attention to the IR region, one can consider the
dependence on the number of branes. As in all other systems, $N_2$
must be large to obtain a reliable classical geometric
description. Then, changing $N_6$ (relative to $N_2$), the theory
goes through several different phases. For small $N_6$, the weakly
coupled description is 11-dimensional and the geometry is $AdS_4$
times the orbifold $S_7/\ZZ_{N_6}$. This system was studied in
\cite{FKPZ9803,Gomis9803,EntGom}.
By increasing $N_6/N_2$ one goes to a 10-dimensional phase. The
geometry still contains an $AdS_4$ part, but now it is {\it
fibered} over a 6-dimensional compact base manifold $X_6$. This is
called {\em a warped product} $AdS_4 \times_w X_6$.
It was shown some time ago \cite{Nieuwe83} that the warped product 
$AdS_{n+1} \times_w X_m$ is the most general metric with an $SO(2,n)$
isometry. Since this isometry is required for a relation to an
$n$-dimensional conformal field theory, the exploration of such spaces is
important in the study
of conformal field theories using Maldacena's duality%
\footnote{Warped products were also considered, for example, in
\cite{dWNi}-\cite{KhPiWa}. Among those discussed so far in the context of
the AdS-CFT duality, the one that was constructed explicitly
\cite{GRW86},\cite{GPPZ},\cite{DZ10206} does not preserve any
supersymmetry. Moreover, it was shown \cite{DZ10206} to be
unstable and, therefore, unlike the geometries studied in the
present work, is not expected to be dual to a conformal field
theory. \ifinter\nl[An instability presumably corresponds to
non-unitarity \cite{KhPiWa}.]\fi}.

When $N_6 \gg N_2$ the 10-dimensional geometry becomes
very highly curved, so we expect that, as in \cite{IMSY9802}, this signals a
transition towards a weakly coupled phase of the gauge theory.
We check this possibility using field-theoretical considerations
and present some evidence that for $N_6 \gg N_2 \gg1$,  the
theory indeed simplifies dramatically and may very-well be weakly coupled.

So far we described the phase structure when the D2 branes are
{\em on} the D6 branes. Moving them off the D6 branes corresponds
to a non-zero mass $m_Q$ for the quarks. We analyze the
corresponding geometry as well, and show that it describes
correctly the renormalization group flow, from the conformal fixed
point with $m_Q=0$, as described above, to another fixed point in
which the quarks decouple. This second fixed point is the IR limit
of pure $N=8$ SYM theory (the world-volume theory of D2 branes
with no other branes), which is dual to M theory on $AdS_4 \times
S^7$ -- the near horizon geometry of M2 branes. For large $N_6$,
there is also an intermediate phase, where the M2 branes are
smeared over the 11th dimension.

In the above description of the phase structure, we implicitly assumed
specific relations between the various parameters of the system, to display
all the possible phases. This includes
\[ g_s \ll N_2 g_s \ll 1 \hs. \]
We also assumed that the mass $m_Q$ of the
quarks is smaller then the scale $g_\YM^2$ set by the gauge coupling.
Now one can increase the mass. Pictorially, this can be done by moving
the horizontal axis in figure \ref{f-phs} up. The right-hand side of the
figure will then deform smoothly to become as the left-hand side,
reflecting the decoupling of the quarks at an increasingly higher energy.
In this context, observe that the order of phases is the same on both
sides: localized M2 branes, smeared M2 branes, D2 branes, field theory and,
finally, the full string theory.

All the information that we obtain from the geometric description,
including the phase structure described above, is in agreement
with known facts about the gauge theory and also with many
expectations. This is a further strong support for the duality
conjecture. One of the old expectations is that,  while a gauge
theory with fields only in the adjoint representation should be
described, in the large $N$ limit, by {\em closed} strings, the
introduction of fields in the fundamental representation should
correspond to the appearance of {\em open} strings. This is indeed
realized in the present system: the D2 branes are replaced
completely by their near-horizon geometry so, without other D
branes, the theory is that of closed strings. On the other hand,
the D6 branes do not disappear and a corresponding singularity in
the geometry is a sign that the open strings ending on the D6
branes exist as additional degrees of freedom.

The structure of this work is as follows. In section \ref{sec-FT}
we briefly review some relevant information about the gauge
theory. In section \ref{s-D2-0} we analyze the geometry of D2
branes located on D6 branes. We determine the phase structure and
consider other issues that relate the geometry and the gauge
theory: the UV-IR relation and the potential between quarks. The
phase structure is then extended, in section \ref{s-D2-Q}, to D2
branes located off the D6 branes. In section \ref{s-dual} we
consider the qualitative changes that occur when some of the
directions in the geometry are compactified, and the corresponding
relations to other systems. In particular, it is explained why in
the D1-D5 system, unlike the present one, the geometric
description is that of D1 branes {\em smeared} over the D5 branes.
Appendix \ref{sec-D6} reviews the construction of the geometry
near D6 branes. Finally, in Appendix \ref{sec-NS5} we briefly
analyze the system of NS5 branes located on D6 branes and obtain a
phase structure very similar to that of the D2-D6 system.

\vspace{1cm}
\noindent{\bf Note Added:}

After the completion of this work, the paper \cite{MP9903} appeared,
where the question of localization of branes was considered.
In accordance with our suggestion here (in section \ref{s-D2-Q}),
it is shown there that smearing (delocalization) of branes in a
classical supergravity solution
corresponds in a dual field-theoretical description to quantum
tunneling between classical vacua \cite{MW66}\cite{coleman}.
This, in particular, confirms our identification of the geometry
of {\em localized} D2 branes as the one relevant for the description of
the dynamics of the D2 branes.

\newsection{The Field Theory on the D2 Branes}
\secl{sec-FT}

We will consider $N_2$ D2 branes, extended in the $x_0,x_1,x_2$
directions.
In the absence of other branes, the low energy effective field
theory on the D2 branes is 3D $N=8$ SYM theory with gauge group
$U(N_2)$.
\internote{This is a dimensional reduction of 10D $N=(1,1)$ SYM.}
The fields are related to fundamental strings stretched
among the D2 branes. The bosonic fields are a gauge field and 7
real scalar fields $X_3-X_9$, all transforming in the adjoint
representation of $U(N_2)$. The scalars describe transverse
fluctuations of the D2 branes and (the eigenvalues of) their vev's
parametrize the moduli space of vacua.

By adding $N_6$ D6 branes, extended in the $x_0-x_6$ directions,
one breaks half of the supersymmetries, leaving $N=4$
supersymmetry on the D2 branes. The original $N=8$ gauge multiplet
decomposes into an $N=4$ vector multiplet plus an adjoint
hypermultiplet. In addition, one generates a set of hypermultiplets
-- ``quarks'' -- that originate from strings stretched between a
D6 brane and a D2 brane. These hypermultiplets, which we denote
generically by $Q$, transform in the fundamental representation
of $U(N_2)$ and also in the fundamental representation of a global
$U(N_6)$ symmetry. Each of them contains 4 scalars.

\subsection{The Moduli Space of Vacua}

Classically, the moduli space of vacua decomposes into branches,
each being a product of two factors: a ``Coulomb'' factor and a
``Higgs'' factor. The ``Coulomb'' factor is parametrized by
scalars belonging to vector multiplets: $X_7-X_9$ and the scalar
$X_\#$ which is the dual of the gauge field; the ``Higgs'' factor
is parametrized by scalars belonging to hypermultiplets: $X_3-X_6$
and $Q$. At a generic point in the moduli space, most of the gauge
symmetry is broken and the remaining massless excitations are
free.
The origin of the moduli space is a singular point, where all the branches
meet%
\footnote{This is when all the fundamental hypermultiplets have the same
mass, as is the case here, since we consider coinciding D6 branes. The
origin of the moduli space corresponds to coinciding D2 branes,
embedded in the D6 branes.}
and the full gauge symmetry is restored. The gauge coupling
constant has a positive mass dimension, therefore the effective
(dimensionless) coupling vanishes in the UV (asymptotic freedom)
and diverges at low energy, apparently leading to a non-trivial IR
fixed point at the origin.

This situation could be modified by quantum corrections, but
non-renormalization theorems impose severe restrictions on such
corrections (as explained, for example, in \cite{InSe9607}).
In particular, the singularity at the origin of the moduli space is not
resolved \cite{InSe9607,mirror}.
\internote{A geometrical argument is given in the supplements.} 
For the brane configuration, this means that the gauge
theory on the coinciding D2 branes flows in the IR to an
interacting conformal field theory. This is the theory which is
conjectured to be dual to string theory in the geometry described
in section \ref{s-D2-0}.

\subsection{Large Number of Flavors}
\secl{s-largeN}

In subsection \ref{s-phase-0}, we will be led to expect that for $N_6\gg N_2$,
the field theory has a weakly coupled description.
In this subsection we will provide support for this possibility,
from field-theoretical considerations.
The relevant limit is%
\beql{largeN} N_2\goto\infty \hsc g^2
N_2=\mbox{const.} \hsc \nu \equiv N_6/N_2 = \mbox{const.} \hs,
\eeq
with large $\nu$.%
\footnote{Note that in 3 dimensions, the gauge coupling $g_\YM$ is not
dimensionless. The parameter $g$ in (\ref{largeN}) is
the dimensionless ratio $g=g_\YM/\sqrt{E}$, where $E$ is the characteristic
energy scale of the process considered.}
$\!\! {}^,$\footnote{Unlike in 4 dimensions, here the number of flavors
is not restricted by the requirement of asymptotic freedom, so $\nu$ can
indeed be large.}
\internote{The large $\nu$ limit was discussed in 4 dimensions with a lattice
regularization. See, for example [Banks\&Zaks82].}
As was shown in \cite{tHooft74}\cite{Venez76}, in the limit
(\ref{largeN}) the theory simplifies and only a small portion of
the Feynman diagrams contributes. Specifically, a Feynman diagram
can be shown to define a surface and the leading contribution to a
given correlation function comes only from ``planar'' diagrams --
those that define surfaces with the topology of a sphere.
Moreover, considering correlation functions of gauge-invariant operators,
one finds that all the $n$-point functions with $n>2$ vanish in the limit
(\ref{largeN}). In non-supersymmetric QCD, such a behavior, together with
the assumption of confinement, was used to argue that in the limit
(\ref{largeN}), the theory is indeed free, the interaction being suppressed by
powers of $1/\sqrt{N}$ (see \cite{Witten79} and references therein).

In the present model we do not expect confinement and, in fact,
the theory is not expected to be free for generic $\nu=N_6/N_2$
(since, for small $\nu$, we will find a weakly coupled geometric dual).
One should, therefore, look for something special that happens only for
large $\nu$.
In fact, a further simplification does occur in this situation: the
leading diagrams are restricted not only to be planar but also to have only
``quark loops'', meaning that if the quark loops are contracted to points,
one should obtain a tree diagram%
\ftl{nu-exp}{All other planar diagrams are suppressed by
powers of $1/\nu$.}.
This is illustrated in figure \ref{f-dig}, where solid lines
represent propagators of fields in the fundamental representation
of the gauge group (quarks), the dashed ones represent propagators
of fields in the adjoint representation and the stars ``*''
represent external vertices (insertions of gauge invariant operators).
\begin{figure}
\pct{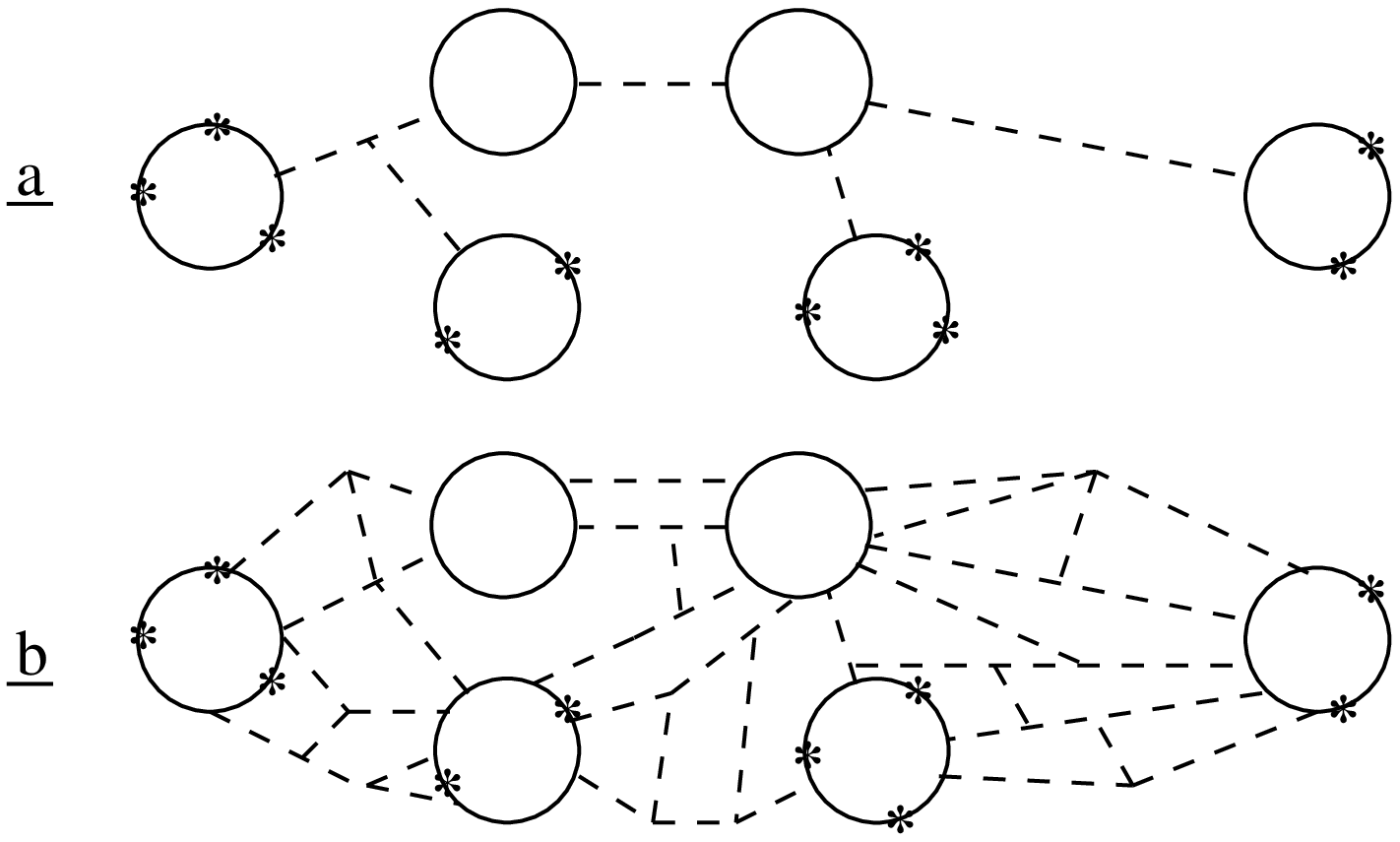}
\capl{f-dig}{Leading diagrams in the large $N$ limit. (a)
large $\nu$; (b) generic $\nu$.}
\end{figure}
The structure of such diagrams is quite simple.
In fact, the sum over all possible internal quark loops can be performed
exactly (it is essentially a geometric sum), leading to a finite number of
effective tree diagrams.
\internote{ \nlb There is a finite number of diagrams with no
internal quark loops \nl (note , however, that they can differ in
number of external loops) \nlb All the diagrams with internal
loops can be obtained from those without them by ``blowing up''
flavor loops. \nlb Unlike color loops, The flavor loops cannot
touch each other (since there are no fields in the bi-fundamental
of $SU(N_f)$). \nlb The internal quark loops can, therefore, be
``absorbed'' into a ``dressed gluon propagator''. \nlb although
the analysis is performed at weak coupling, aposteriory it could
be possible to extrapolate the (exact) result to strong coupling.
and, presumably, the $\nu$ dependence would be the same.}
This simplification is characteristic to the large $N$ limit of
vector models: models with a global symmetry group $U(N)$, $SO(N)$ or
$USp(N)$, in which all the fields are in the vectorial representation of the
symmetry (or the trivial one). It typically leads to the conclusion that the
theory is free in this limit%
\footnote{One such example is the Gross-Neveu model \cite{GN74}.}.
Presumably, this is what happens also in the
present case, although we will not try to verify this here.

\newsection{D2 Branes Localized On D6 Branes}
\secl{s-D2-0}

{}From now on, we turn our attention to the geometry of the D2-D6
system, and investigate its relation to the 3 dimensional gauge
theory.

In this section we consider D2 branes localized on the D6 branes.
This corresponds, in the field theory, to a vanishing mass for the
fundamental hypermultiplets $Q$.
We determine the corresponding geometry in 11 and
10 dimension, analyze the range of validity of the descriptions
obtained and discuss the phase structure that emerges. We also estimate the
entropy and show that it is consistent with smooth transitions between the
field-theoretic and geometric phases.
Then we consider other issues that relate the geometry and the gauge
theory: the UV-IR relation and the potential between ``quarks''.

\subsection{The Geometry}
\secl{geom}

There are several ``D2-D6'' geometries that one could try to relate to
the field theory described in the previous section and, therefore,
it is important to determine which is the correct one.
Probably the most familiar one is the configuration in which the
D2 branes are smeared along the directions of the D6 branes.
However, the corresponding supergravity solution \cite{PT}-\cite{AEH}
is {\it not} the relevant one in the present context\footnote{For
reviews of (intersecting) brane solutions in supergravity, see
\cite{p-branes}-\cite{Skend}}. One indication
for this is that the near horizon geometry does not have the $SO(2,3)$
isometry group, necessary for a relation to a 3 dimensional conformal
field theory. This issue will be discussed further in section \ref{s-dual},
where it will become clear that the correct configuration is that in
which the D2 branes are fully localized inside the D6 branes
(and of course also in the overall transverse space).
The corresponding supergravity solution near the horizon of the D6 branes
has been presented in \cite{ITY3103} and can be
understood by noting the following facts%
\footnote{In order to make our paper self-contained, and also in
order to establish our notations, these facts are reviewed in
Appendix \ref{sec-D6}}:
A set of $N_6$ coinciding D6 branes corresponds in M theory to a
Kaluza-Klein (KK) monopole \cite{Town}.
The geometry is that of a $\RR^{6+1}\times\mbox{Taub-NUT}$ space
and, close to the center, it is well-approximated
by the orbifold $\RR^{6+1}\times\RR^4/\ZZ_{N_6}$.
D2 branes correspond in M theory to M2 branes, so to add them, one starts
from the flat covering space of the orbifold,
in which the M2 brane solution is well-known, and
then makes the orbifold identifications.

In the covering space of the orbifold, there are $N_6N_2$ images of M2
branes.
When they all coincide (at the singularity), the corresponding metric
for {\em extremal} branes is \cite{M2}
\beql{s-M2}
ds_{11}^2=f_2^{-\frac{2}{3}}dx_\prl^2
+f_2^{\frac{1}{3}}(dr^2+r^2d\Om_7^2) \hs,
\eeq
where
\beql{f2-def} f_2=1+\frac{32\pi^2N_6N_2l_p^6}{r^6} \hsc
 x_\prl = \{ x_0, x_1, x_2 \} \hsc \eeq
\internote{$A^{(3)}=\frac{f_2-1}{f_2}dx^0\wedge dx^1\wedge dx^2$.}
and $d\Om_7^2$ denotes the metric on the unit 7-sphere. We split
the directions transverse to the M2 branes into directions
parallel to the singularity (along which in 10 dimensions the D6
branes will be extended) and directions transverse to it. This
leads to the following parametrization of the 7-sphere:
\beql{Om7-def} d\Om_7^2=d\bt^2+\cos^2\bt d\Om_{3\prl}^2+\sin^2\bt
d\Om_{3\perp}^2 \hs. \eeq The metrics $d\Om_{3\prl}^2$,
$d\Om_{3\perp}^2$ describe two unit 3-spheres
$S^3_\prl,S^3_\perp$, while the additional angle
$0\le\bt\le\frac{\pi}{2}$ measures the orientation with respect to
the singularity ($\bt = 0$ denoting points at the singularity).

The orbifold projection acts
only on $S^3_\perp$, so the orbifold metric is obtained by replacing
$d\Om_{3\perp}^2$ by the metric $d\tilde\Om_{3}^2$ on
$S^3_\perp/\ZZ_{N_6}$:
\beql{s-s3}
d\Om_{3\perp}^2 \goto d\tilde\Om_{3}^2  = \frac{1}{4} d\Om_2^2
+ \left[\frac{1}{N_6} d\psi + \frac{1}{2}(1-\cos\th) d\vph\right]^2 \hs,
\eeq
\[ d\Om_2^2=d\th^2+\sin^2\th d\vph^2 \hs, \]
where $0\le\th\le\pi$ and $\vph,\psi$ are periodic with period $2\pi$.%
\footnote{As described in Appendix \ref{sec-D6}, the expression
$\frac{1}{4} d\Om_2^2 +
\left[\frac{1}{N_6} d\psi + \frac{1}{2}(1-\cos\th) d\vph\right]^2$
parametrizes the unit 3-sphere when $\psi$ ranges from $0$ to
$2\pi N_6$. The $\ZZ_{N_6}$  action leading to an orbifold is given
by $\psi \mapsto \psi + 2\pi$.}

Next we reduce to 10 dimensions, identifying $\psi$ as the coordinate of
the compact 11th dimension \cite{ITY3103}.
The 10 dimensional metric $ds_{10}^2$ (in the string frame)
and the dilaton $\phi$ are identified through
\beql{11-10} ds_{11}^2=e^{4(\phi-\phi_\infty)/3}(R_\#d\psi+A_\mu dx^\mu)^2
            +e^{-2(\phi-\phi_\infty)/3}ds_{10}^2 \hs,
            \eeq
where $R_\#$ is the asymptotic radius of the 11th dimension. It is related to
the string scale $l_s$ and string coupling $g_s=e^{\phi_\infty}$ by
$R_\#=g_sl_s$
(see eq. (\ref{lgs})). This gives
\beql{s-D2} ds_{10}^2=e^{2(\phi-\phi_\infty)/3}
\{f_2^{-\frac{2}{3}}dx_\prl^2
+f_2^{\frac{1}{3}}(dr^2+r^2d \tilde\Om_6^2)\} \hs,
\eeq
\[ d \tilde\Om_6^2 = d\bt^2+\cos^2\bt d\Om_{3\prl}^2
                 +\frac{1}{4}\sin^2\bt d\Om_2^2 \]
\internote{The angular space is obtained from $S^6$ by shrinking $S^2$ by
a factor of 2 $\goto$ conical singularity at $\bt=0$.}
and
\beql{phi-D2}
e^\phi=g_sf_2^{\frac{1}{4}}
\left(\frac{r\sin\bt}{N_6 g_s l_s}\right)^{\frac{3}{2}} \hs.
\eeq
\internote{$e^\phi = g_s f_2^{\frac{1}{4}} f_6^{-\frac{3}{4}}$
(also for $r_Q>0$ below).}

\newpar{The Near Horizon Geometry}

Near the M2 brane horizon (when $f_2\gg1$), the metric
(\ref{s-M2}) simplifies to
\beql{s-M2-n} ds_{11}^2 = l_p^2
\left[\frac{U^2}{(32\pi^2 N_6 N_2)^\frac{2}{3} } dx_\prl^2
          + (32\pi^2 N_6 N_2)^\frac{1}{3}
          \left(\frac{dU^2}{4U^2} + d\tilde\Om_7^2
          \right)\right] \hs,
\eeq
where\footnote{In the Maldacena or decoupling limit
$l_p \rightarrow 0$, the coordinate $U$ must be kept fixed in order to
obtain a well-defined supergravity action \cite{Mald9711}.}
\beql{U-def-M2}  U=\frac{r^2}{l_p^3}
\eeq
and $d \tilde\Om_7^2$ stands for the metric on $S^7/ \ZZ_{N_6}$, as
described above. The metric (\ref{s-M2-n}) describes the product-space
$AdS_4 \times S^7/\ZZ_{N_6}$, where the radius of the $AdS_4$ is given by
\beql{R11}
R_{\rm AdS}^{(11)} = l_p \left( \frac{\pi^2 N_6 N_2}{2} \right)^\frac{1}{6}
\hs.
\eeq
The relation between M-theory on this $AdS_4 \times S^7/\ZZ_{N_6}$
space and three dimensional field theory has been considered in
\cite{FKPZ9803}\cite{Gomis9803}\cite{EntGom}.

The 10D geometry becomes
\beql{s-D2-n}
ds_{10}^2 = l_s^2 \, \frac{\sin\bt}{N_6}
\left[\frac{U^2}{\sqrt{32\pi^2 N_6 N_2}} dx_\prl^2
+ \sqrt{32\pi^2 N_6 N_2}
\left(\frac{dU^2}{4U^2}+d \tilde\Om_6^2 \right)\right] \hs,
\eeq
\beql{phi-D2-n}
e^\phi = \left( \frac{32\pi^2 N_2}{{N_6}^5}\right)^\frac{1}{4}
         (\sin\bt)^\frac{3}{2} \hs,
\eeq
where $d\tilde\Om_6^2$ is as in (\ref{s-D2}).
Note that $(x_\prl,U)$ still parametrize an $AdS_4$ space and the
dilaton $\phi$ is independent of these coordinates, so this configuration
has an $SO(2,3)$ symmetry.
However, the radius of the $AdS_4$ space (as well as the dilaton)
{\it depends on the orientation} (the angle $\bt$):
\beql{R-AdS-D2}
R_{\rm AdS}^{(10)} = l_s \left( \frac{\pi^2 N_2}{2 N_6} \right)^\frac{1}{4}
\sqrt{\sin\bt} \hs.
\eeq
Therefore the geometry is that of $AdS_4$ {\em fibered} over a compact
manifold $X_6$
(parametrized by the coordinates of $d\tilde\Om_6^2$)%
\footnote{As reviewed in the introduction, such a geometry is also called  a
``warped'' product.}.

As already mentioned at the beginning of this subsection, the
emergence of a geometry with $SO(2,3)$ symmetry (which can,
therefore, be dual to a conformal field theory) depends crucially
on the fact that the D2 branes are {\em localized} on the D6
branes. Indeed, for D2 branes smeared in $k$ (compact) dimensions
of the D6 branes ($1 \le k \le 4$), the corresponding harmonic
function is $f_2\sim r^{6-k}$ and the dilaton is no longer
independent of $r$ (see eq. (\ref{phi-D2})). This will be
discussed further in section \ref{s-dual}.

Another symmetry of the theory on the branes that is realized geometrically is
the R-symmetry. These are space-time rotations that leave the M2 brane
invariant. In the absence of D6 branes, the R-symmetry is the $Spin(8)$
symmetry of the 7-sphere in eq. (\ref{s-M2}).
\internote{Remarks:
\nlb Only a $Spin(7)$ subgroup is preserved by the SYM action at finite
coupling. -- this is the group of $x^3-x^9$ rotations, so the gauge field is
invariant under it and the scalars transform in the vectorial representation.
\nlb The gauge field in three dimensions is dual to a compact scalar $X_\#$.
In the limit of infinite gauge coupling (\eg\ the IR limit), the scalar
$X_\#$ decompactifies and full $SO(8)$ R-symmetry is recovered.}
\internote{An apparent puzzle:
\nlb Translations of $X_\#$ seem as another (global) $U(1)$ symmetry of the
gauge theory but the zero mode of $x_\#$ is not a physical degree of freedom!
(only $\pt_\mu X_\#$ appears in the duality relation to the gauge field).
\nlb In the geometric description, these translations do not leave the
M2 brane invariant, so they indeed should not be realized on the brane!
\nlb This, however, changes when the D2 brane is on D6 branes! }
With D6 branes, the 3D supersymmetry is reduced to $N=4$ and, correspondingly,
the R-symmetry is reduced to $Spin(4)$.
In the present description, the D6 branes are represented by the $\ZZ_{N_6}$
orbifold and it breaks the $Spin(8)$ symmetry as follows:
the orbifold plane breaks $Spin(8)$ to $Spin(4)_\prl\otimes Spin(4)_\perp$,
acting on the two 3-spheres in eq. (\ref{Om7-def}).
Decomposing further: $Spin(4) \equiv SU(2)_L \otimes SU(2)_R$,
$\ZZ_{N_6}$ is a subgroup of one of the $SU(2)$ factors of $Spin(4)_\perp$,
say $SU(2)_{\perp R}$, and it breaks this factor to $U(1)_{\perp R}$.
Comparing to eq. (\ref{s-s3}), $SU(2)_{\perp L}$ is the symmetry of the
2-sphere parametrized by $\th$ and $\vph$ and $U(1)_{\perp R}$ acts as
translations in $\psi$.
Now, by analyzing the transformations of the various fields in the field
theory under $SU(2)_{\prl L} \otimes SU(2)_{\prl R}
\otimes SU(2)_{\perp L} \otimes U(1)_{\perp R}$, 
one identifies $SU(2)_{\prl R}\otimes SU(2)_{\perp L}$ as the R-symmetry
group, while $SU(2)_{\prl L} \otimes U(1)_{\perp R}$ is a global symmetry.
\internote{Remarks:
\nlb The scalar fields transform under
$SU(2)_{\prl L} \otimes SU(2)_{\prl R}
\otimes SU(2)_{\perp L} \otimes U(1)_{\perp R}$
in the following representations:
$X_\#$ in $(1,1,1)_1$, $X_3-X_6$ in $(2,2,1)_0$, $X_7-X_9$ in
$(1,1,3)_0$ and $Q_i$ in $2(1,2,1)_0$.
\nlb What is the geometrical meaning of the fact that the $Q$'s are in
a spinorial representation of $Spin(4)_\prl$?
\nlb The action of $SU(2)_{\prl L}$ on the $X$'s follows from that of
$Spin(7)$ and it can be chosen to be trivial on the $Q$'s, since they couple
only to those $X$'s that are $SU(2)_{\prl L}$-invariant.
\nlb It is useful to view the theory as a dimensional
reduction from six dimensions. In the absence of the D6 branes, there is
$N=(1,1)$ supersymmetry and $Spin(4)_\prl=Usp(2)_L \otimes Usp(2)_R$ is the
R-symmetry.}

\subsection{The Phase Structure}
\secl{s-phase-0}

In this subsection we consider the near-horizon geometries of section
(\ref{geom}) and analyze the changes in the description of the corresponding
theory as the parameters are varied.
We will see, in particular, that the description
changes as we vary the number of flavors relative to the number of
colors.

\newpar{The Classical Approximation and
The Dependence on the Number of Branes}

We obtained, in the previous subsection, two geometric descriptions -- 10
and 11 dimensional -- of which only in one the quantum corrections can be
expected to be small. The transition between 11 dimensional and 10 dimensional
geometries is governed by the radius $R_\psi$ of the circle parametrized by
$\psi$ or, equivalently, by the dilaton $\phi$ given in eq. (\ref{phi-D2-n}):
\[ \frac{R_\psi}{l_p} = e^{2\phi/3}
= \left(\frac{32\pi^2N_2}{N_6^5}\right)^\frac{1}{6} \sin\bt \hs, \]
\internote{$R_\psi = \frac{1}{N_6} f_2^{\frac{1}{6}} r \sin\bt$
(also for $m_Q>0$).}
so the geometry is 10 dimensional iff%
\footnote{Actually, the circle $S^1$ parametrized by $\psi$ is
contractible (at the orbifold singularity $\bt=0$), so one might
worry that, like the polar angle in the plane, it cannot be
identified as a small compact dimension. To exclude such a
possibility, one could require that the radius of the circle
$R_\psi$ is much smaller than the distance $\delta$ from the
contraction point. However, whenever $R_\psi/l_p$ is small, this
condition is automatically satisfied almost everywhere, i.e.
except at distances from the contraction point which are much
smaller than $l_p$. In the present case it is easy to see that
$R_\psi / \delta = 1/N_6$ so this is indeed small, due to
(\ref{cond}).}
\beql{cond}
N_2 \ll N_6^5 \hs.
\eeq
\internote{Otherwise, $R_\psi$ is small only near the singularity}

Another condition for the validity of the classical geometric description is
small curvature.
In the 11 dimensional geometry (\ref{s-M2-n}), the curvature ${\cal R}^{(11)}$
is
\[ l_p^2 {\cal R}^{(11)} \sim (N_6 N_2)^{-\frac{1}{3}} \hs, \]
so the classical description is reliable for large $N_6 N_2$.
In the 10 dimensional geometry (\ref{s-D2-n}), we obtain
\beql{curv} l_s^2 {\cal R}^{(10)} \sim
\sqrt{\frac{N_6}{N_2}}\frac{1}{\sin^3\bt} \eeq
\internote{$l_s^2 R = \frac{1}{a_0^2\sin\bt}(34-21\cot^2\bt)$,
$a_0^4=\frac{8\pi^2 N_2}{N_6}$.} and it is small away from the D6
branes iff $N_6 \ll N_2$.
At the location $\bt=0$ of the $D_6$ branes, the curvature diverges and this
singularity is identified as a sign that, in this geometric description, some
degrees of freedom related to the D6 branes were effectively integrated out.
Therefore, this singularity is expected to be resolved by adding these
degrees of freedom%
\footnote{The 11 dimensional geometry is also singular at $\bt=0$
(although the singularity is milder -- an orbifold singularity)
and this is understood in the same way as in 10 dimensions. See
also \cite{AFM6159} for a similar situation.}.
It is clear what these degrees of freedom are: they are the open strings
ending on the D6 branes. This is a realization of the original
expectations about the relation between strings and gauge theory, namely,
that pure Yang-Mills theory should be described by {\em closed} strings,
while adding quarks corresponds to adding {\em open} strings.
Indeed, here the pure SYM theory (with $N_6=0$), is conjectured to be dual
to type IIA string theory in the background of the near-horizon geometry of
D2 branes \cite{IMSY9802}, which is a theory of closed strings%
\footnote{Although, at low energies,
the coupling becomes large and the more appropriate description is M theory.},
while quarks appear as open strings ending on D6 branes.
Observe that the D2 and D6 branes behave differently in this
context: while the D2 branes ``disappeared''
and are fully represented by the geometry, the D6 branes remain as
D branes, {\em in addition} to their influence on the geometry.

To summarize, to obtain a classical geometric description (\ie, with small
curvature), $N_2$ must be large.
When this is satisfied, there are three phases, depending on
the relation between $N_2$ and $N_6$.
For $N_2^{\frac{1}{5}}\ll N_6\ll N_2$ there is a geometric
description in 10 dimensions, for smaller $N_6$ there is such a description
in 11 dimensions, and for larger $N_6$ there is no geometric description,
as the curvature becomes large.
In similar situations considered in the literature
(\eg, refs. \cite{IMSY9802} \cite{AFM6159}),
in the region of large curvature the field theory became weakly coupled.
As was discussed in subsection \ref{s-largeN}, it is plausible that this is
what happens also in the present model and for $N_2\ll N_6$,
there is a weakly coupled field-theoretical description.

It is interesting to see what the two types of quantum corrections in the
10 dimensional description correspond to in the field theory.
The string loop expansion is in $e^\phi$, holding the geometry fixed.
Holding the metric (\ref{s-D2-n}) fixed means
\[ \nu \equiv \frac{N_6}{N_2}=\mbox{const.} \]
\internote{Details:
\nl The metric is
\nl $ds_{10}^2 = l_s^2 \sin\bt
\left[ \left(\frac{V'}{\wbar{V'}}\right)^{\frac{1}{2}} V'^2 dx_\prl^2
+ \frac{a^3}{N_6} \left(\frac{V'}{\wbar{V}'}\right)^{\frac{3}{2}}
\left(\frac{d{V'}^2}{4{V'}^2}+d\tilde\Om_6^2 \right)\right]$,
\nl where ${V'}^2=U^2/a^3 N_6 \sim \sqrt\nu V^2$, $V=U/N_6$
\nl so the quantity that must be kept fixed are
\nl $\nu$ ($\sim\frac{N_6^2}{a^6}\sim(\frac{ds^2}{l_s^2})^{-2}$);
\nl in the massive cases: also $m_Q$
(since $\nu m_Q^4 \sim \nu \left(\frac{U_Q}{N_6}\right)^4$).
\nl in the non-extremal case, also $\frac{\ep}{N_6^3}$
(since $\nu \left(\frac{\ep}{N_6^3}\right)^\frac{4}{3}
\sim \nu \left(\frac{U_h}{N_6}\right)^4 \sim V_h^4)$.}
and, using eq. (\ref{phi-D2-n}), we see that the expansion is in small
\[ e^\phi \sim \left(\frac{N_2}{N_6^5}\right)^{\frac{1}{4}} \sim \frac{1}{N_2}
\hs. \]
This is precisely the large $N$ expansion considered in subsection
\ref{s-largeN} (see eq. (\ref{largeN})).
Now, from eq. (\ref{curv}) we see that the curvature ($\al'$)
corrections of string theory lead to an expansion in small $\nu$.
This is also in agreement with the field theory, as described in
subsection \ref{s-largeN}: there, the sub-leading diagrams are
suppressed by $1/\nu$ (see footnote \ref{nu-exp}), so $\nu$ indeed
controls the transition between the two descriptions.
Finally, we note that $1/N_6$ has in these expansions the same
role as $g_\YM^2$ in the consideration of D3 branes (where the
expansion parameters are $1/N$ and $1/g_\YM^2N$ respectively, with
$N$ being the number of colors, as $N_2$).

\newpar{The Near-Horizon Range}

Next, we analyze the implications of the fact that we consider
only the near-horizon region. The near-horizon region of the D2
branes is characterized by the requirement $f_2\gg1$. Using the
eqs. (\ref{f2-def}),(\ref{U-def-M2}) and (\ref{lgs}) this
translates into the requirement \beql{rel1} U^3
\ll\frac{N_6N_2}{g_\YM^2 l_s^4} \hs, \eeq
where
\beql{gYM}
g_\YM^2= \frac{g_s}{l_s} = \frac{R_\#^2}{l_p^3} \eeq
is the gauge
coupling constant in the field theory on the D2 branes. This is
the condition for the decoupling of the branes from the bulk and
it can be satisfied for any given $U$,
by taking $l_s$ small enough, while keeping $g_{YM}^2$ fixed.

The near-horizon region of the D6 branes is characterized by
$f_6\gg1$, where $f_6$ is defined in eq. (\ref{f6-def}). By
identifying $r^2 \sin^2 \bt = l_p^3 U \sin^2 \bt$ with $\rho^2$ in
eq. (\ref{rho-def}) and, using the relations (\ref{lgs}),
this requirement can be written as \beql{f6-large}
U\ll\left(\frac{N_6}{\sin\bt}\right)^2 g_\YM^2 \hs. \eeq
\internote{Details:
$1 \ll f_6 = 1 + \frac{N_6 R_\#}{2r_6}
= 1 + \left(\frac{N_6 R_\#}{r \sin\bt}\right)^2$}
This restriction is non-trivial also for $l_s\goto0$, so it should be
understood in the framework of the dynamics of D2 branes {\em decoupled
from the bulk}, \ie, in the framework of the gauge theory.
As will be explained below, the coordinate $U$ is proportional to the
energy scale in the field theory, so the restriction (\ref{f6-large})
means that the geometric description we consider corresponds in the
gauge theory only to energy scales which are low compared to a scale $E_{IR}$
set by the gauge coupling $g_\YM$ (and $N_2$, $N_6$).
In this context, note that the parameters $g_s$ and $R_\#$ do not appear in
the geometry described in the previous section,
so the geometry is independent of $g_\YM$, as expected for energy scales
far below the scale set by $g_\YM$.

Going further from the D6 branes, their influence on the geometry becomes
negligible and the geometry is approximately that of D2 branes in isolation.
This system was analyzed in \cite{IMSY9802}. The geometry is no longer
$SO(3,2)$-invariant, so this region corresponds, in the gauge theory,
to energy scales above those governed by the IR fixed point.
At a sufficiently high energy -- in the range of the UV fixed point --
the field theoretic description becomes weakly coupled%
\footnote{At the corresponding distances from the D2 branes, the curvature
is large and the classical geometrical description is not reliable.}.
The intermediate region, in which the gauge theory is strongly coupled but
not conformally invariant, is bounded from above by
\[ E_{UV}\sim N_2 g_\YM^2 \hs, \]
(since the dimensionless parameter $g_\eff$ controlling the quantum
corrections in the gauge theory
is $g_\eff^2 = \frac{N_2 g_\YM^2}{E}$).
To determine the lower bound $E_{IR}$, as implied by eq.
(\ref{f6-large}), we need the geometric identification of the
field-theoretic energy scale $E$. This will be discussed in some
detail below (in subsections \ref{s-en} and \ref{s-phase-Q}), but
already at this stage we can make some simple observations.
First, realizing that $U$ is the only dimensionfull quantity in the geometry
(\ref{s-D2-n}),(\ref{phi-D2-n}),
we deduce that $E$ must be proportional to $U$ (as in any $AdS$ geometry).
Next, consider the large $N$ limit (\ref{largeN}). As discussed above,
the metric (\ref{s-D2-n}) remains fixed in this limit.
On the other hand, the limit is taken with a fixed energy, so the geometric
identification of the energy should depend on $N_2$ and $N_6$ only through
$\nu$. This leads to the general form%
\footnote{This can be seen by observing that when the metric (\ref{s-D2-n})
is written in terms of $V=\frac{U}{N_6}$ instead of $U$, it indeed depends on
$N_2$ and $N_6$ only through $\nu$.}
\[ E \sim \frac{U}{N_6}\nu^\gm \hs, \]
so the lower bound implied by eq. (\ref{f6-large}) is
\[ E_{\rm IR} \sim \nu^{\gm+1} N_2 g_\YM^2 \hs. \]
In the next subsection we will find $\gm=0$ (in processes relevant to
fundamental strings) or $\gm=\half$ (in processes relevant to
thermodynamics and supergravity fields).
For both these identifications one obtains, as required, that
$E_{IR}< E_{UV}$
whenever $\nu\ll1$ (\ie, in the validity range of the classical geometric
description). Note that this is a strict inequality, which means that an
intermediate energy region does exist.
\internote{Details: \nlb Considering the general form
$V=\frac{\nu^\gm}{N_6^\al}U$, the condition $E_{IR} \le E_{UV}$
is: \nl $\frac{E}{U} \sim \frac{\nu^\gm}{N_6^\al} \siml
\frac{1}{\nu N_6}$ whenever $\nu\ll1$. \nlb It is satisfied iff
$\al\ge1$ and $\gm\ge-1$. \nlb In particular, it is satisfied by
$\frac{E}{U} \sim \frac{1}{N_6}$ or $\frac{1}{\sqrt{N_6 N_2}}$ \nl
but not by $\frac{E}{U} \sim \frac{1}{(N_6 N_2)^\frac{1}{3}}$ or
1.}

Finally we remark that both the restrictions on $U$ discussed above can be
removed by taking first $l_s\goto0$ (the decoupling limit of the brane)
and then $g_\YM\goto\infty$ (the IR limit of the gauge theory).
\internote{Comparing to the $N_6=0$ case:
\nlb There the string coupling diverges as one approaches the D2 brane, so
the IR fixed point corresponds to an 11 dimensional geometry,
which means infinite string coupling, in agreement with a diverging effective
gauge coupling.
\nlb Here the large $N_6$ keeps the string coupling small at the IR
(on the horizon) but, as wee see, the relevant effective gauge coupling is,
nevertheless, still infinite (in the above sense).
\nlb Compare to the discussion of the local gauge coupling at the beginning
of the next section (in an internal note).}

\subsection{The Entropy and (no) Phase Transition}

To check the possibility of a phase boundary between the geometric and
non-geometric phases, we calculate here the entropy of the system, when it
is at a non-vanishing temperature.
For this we should consider non-extremal D2 branes.
The non-extremal version of the metric (\ref{s-M2}) is
\beql{s-M2-h}
ds_{11}^2=f_2^{-\frac{2}{3}}(-f_hdt^2+d\vec{x}_\prl^2)
+f_2^{\frac{1}{3}}(f_h\inv dr^2+r^2d\tilde\Om_7^2) \hs, \eeq where
\beql{fh-def} f_h=1-\frac{192\pi^4l_p^9\ep}{7r^6} \hs. \eeq
The parameter $\ep$ is the M2 brane tension above extremality and it is
interpreted in the field theory as the energy density.
Following the derivation in subsection \ref{geom} we obtain, in the
near-horizon region of the D2 branes, the following string metric
\beql{s-D2-h-n} ds_{10}^2
= l_s^2 \frac{\sin \bt}{N_6} \left[\frac{U^2}{\sqrt{32\pi^2 N_6
N_2}} (-f_h dt^2 + d\vec{x}_\prl^2) + \sqrt{32\pi^2 N_6 N_2}
\left(f_h\inv\frac{dU^2}{4U^2}+d\tilde\Om_6^2\right)\right] \hs,
\eeq
with
\beql{fh-n} f_h=1-\left(\frac{U_h}{U}\right)^3 \hsc
U_h^3=\frac{192\pi^4}{7}\ep \hs.
\eeq
\internote{
For Dp branes, $U_h^{7-p}\sim g_\YM^4\ep$, $g_\YM^2\sim g_s l_s^{3-p}$.}
The dilaton is independent of $\ep$ and, therefore, given by eq.
(\ref{phi-D2-n}).

In the geometric description, the entropy $S$ is related, by the
Bekenstein-Hawking relation, to the area $A$ of the horizon
(in the Einstein metric $ds_{\rm Ein}^2 = e^{(\phi-\phi_\infty)/2}ds_{10}^2$).
The horizon is at $U=U_h$ and this gives, for the density of the entropy,
\beql{S-Bek} s \sim
\frac{1}{V_\prl}\frac{A}{l_{10}^8} \sim (N_6
N_2)^{\frac{1}{3}} U_h^2 \sim (N_6 N_2 \ep^2)^{\frac{1}{3}} \hs,
\eeq
where $V_\prl=\int dx^2_\prl$ is the (spatial) volume of the D2 brane
and $l_{10}=g_s^{\frac{1}{4}}l_s$ is the 10-dimensional Planck length.
On the other hand, when the field theory is weakly coupled, the
entropy is approximately that of an ideal gas. This gives
\cite{KT9604} \beql{S-gas} s \sim (n\ep^2)^\frac{1}{3} \hs, \eeq
where $n$ is the number of degrees of freedom which in the present
case is $n\sim N_2^2 +N_6 N_2$. As discussed in subsection
\ref{s-phase-0}, there are two situations where we expect weakly
coupled field theory. One is $N_2 \ll N_6$. At the transition
point $N_2=N_6$, the two expressions (\ref{S-Bek}) and
(\ref{S-gas}) for the entropy agree and this is consistent with a
smooth transition between the two descriptions (\ie\ with no phase
boundary). The other weakly coupled region is at large energy
scales, where the effective gauge coupling $g_\eff$ is small. For
$N_6\ll N_2$, $n\sim N_2^2 \gg N_6 N_2$ so (\ref{S-Bek}) and
(\ref{S-gas}) do not agree, however this is still consistent with
a smooth transition since, as we saw, these phases are separated
by an intermediate phase.
\internote{A PROBLEM:
\nlb Eq. (\ref{S-gas}) gives, for the temperature,
\nl $\frac{1}{T}=\frac{ds}{d\ep} \sim \frac{(N_6 N_2)^{\frac{1}{3}}}{U_h}$
which does not agree with (\ref{T-Hawk}). Is there a mistake?
\nlb Alternatively, assuming the thermodynamics of $N_6 N_2$ free
degrees of freedom, we obtain
$N_6 N_2 T^2 \sim s\sim a^2 U_h^2$, which leads, again, to the same wrong
expression for $T$.}
\internote{Details: \nlb $S=A/4G_N$, $G_N=8\pi^5 l_{10}^8$ \nlb a
free gas of $n$ d.o.f. in $p+1$ dimensions \cite{KT9604}: \nl
$E\sim nVT^{p+1} \hsc S\sim nVT^p\sim(nVE^p)^{\frac{1}{p+1}}$
$\Goto$ $s \sim (n \ep^p)^{\frac{1}{p+1}}$. \nlb The above
comparisons are related to the correspondence principle
\cit{Horowitz\&Polchinski9612146}; reviews:
\cit{Horowitz-grqc9704072;Peet9712253}.}

\subsection{The UV-IR Relation}
\secl{s-en}

It was suggested in \cite{SW9805} (see also \cite{PP9809}) that a
UV cutoff in the field theory -- a minimal distance $\dl x$ -- is
related to an IR cutoff in the geometric description -- an upper
bound $U_\max$ on the distance from the branes. We will check this
suggestion in the present configuration.

\newpar{The Field Equations}

One of the approaches to a UV-IR relation is to consider solutions
of the classical field equations. For a massless scalar field
$\phi$, the equation is
\beql{eq-phi} \nabla^2\phi=0 \hs, \eeq
where $\nabla^2$ is the scalar Laplacian. For the metric
(\ref{s-D2-n}), the Laplacian is
\beql{Lap} l_s^2 \sqrt{8\pi^2
\frac{N_2}{N_6}} \sin\bt \nabla^2 = \sin\bt \tilde\nabla^2_6 +
\nabla^2_{AdS} + 2(\pt_\bt\ln\sin\bt)\pt_\bt \hs, \eeq where
$\tilde\nabla^2_6$ is the Laplacian of $d \tilde\Om_6^2$ and
$\nabla^2_{AdS}$ is the Laplacian for the $AdS_4$ space with unit radius%
\footnote{with the metric $ds_{AdS}^2= \hat U^2
dx_\prl^2+\frac{d\hat U ^2}{\hat U^2} $.}:
\[ \nabla^2_{AdS}= \hat U^2 \pt_{\hat U}^2 + 4 \hat U \pt_{\hat U}
+ \frac{1}{\hat U^2}\pt^\al\pt_\al \hsc \al=0,1,2 \hsc \hat U
=\frac{U}{\sqrt{8\pi^2 N_6 N_2}} \hs.\] Therefore, substitution of
$\phi=e^{ik\cdot x_\prl}\vph(x_\perp)$ (where $x_\perp$ represents
the coordinates of the space transverse to the D2 branes) in eq.
(\ref{eq-phi}) gives the following functional dependence for the
solution:
\[ \phi=e^{ik\cdot x_\prl}\vph(N_6N_2\frac{k^2}{U^2},\Om) \]
(where $\Om$ represents the angles, including $\bt$).
This suggests \cite{PP9809} the following relation
\beql{UVIR-phi}
\frac{1}{\dl x_\prl}\sim k \sim \frac{U_\max}{\sqrt{N_6N_2}}\Ec_\phi(\bt) \hs.
\eeq
Observe that we included in the above relation a dependence on $\bt$, through
a function $\Ec_\phi$.
Such a dependence cannot be excluded, since the geometry depends on $\bt$.
However, the present approach does not provide information on this dependence
(\ie\ on $\Ec_\phi$), since $\bt$ enters independently in eq. (\ref{eq-phi}).
\internote{The relation (\ref{UVIR-phi}) can be deduced directly from the
metric and, therefore, will emerge in any context where $\dl x_\prl$
enters only through the metric, \eg, other SUGRA equations; geodesic equations;
thermodynamic Al considerations.}

\newpar{Thermodynamics}

Another way to obtain a UV-IR relation is through thermodynamical
considerations. For this we consider the geometry of non-extremal D2 branes,
as derived in the previous subsection.
The Hawking temperature $T$ is given by%
\footnote{It is obtained by considering the Euclidean continuation of the
metric and demanding a periodicity $\frac{1}{T}$ of Euclidean time that
eliminates the conical singularity at the horizon.}
\beql{T-Hawk} T=\frac{1}{2\pi}
\frac{d}{dU}\left.\sqrt{-\frac{G_{tt}}{G_{UU}}}\right|_{U=U_h} \sim
\frac{d}{dU}\left.\left(\frac{U^2 f_h}{\sqrt{N_6
N_2}}\right)\right|_{U=U_h} \sim \frac{U_h}{\sqrt{{N_6}{N_2}}}
\eeq
and, assuming that the temperature is bounded by the UV
cutoff $\frac{1}{\dl x_\prl}$, we obtain
\beql{UVIR-T}
\frac{1}{\dl x_\prl} \sim \frac{U_\max}{\sqrt{{N_6}{N_2}}} \hs.
\eeq
The dependence on $U,N_6,N_2$ is as in eq. (\ref{UVIR-phi}).
Since the horizon is characterized by $U=$const., the present
approach suggests a UV-IR relation which is independent of $\bt$
(\ie, $\Ec=$const. in eq. (\ref{UVIR-phi})).

\newpar{A Fundamental String as a Charge}

Finally, we consider a fundamental string with one end on the D2 branes and
the other extending to infinity.
\internote{
When the D2 branes are replaced by the super-gravity fields they induce,
one obtains a fundamental string in this background, ending on the horizon.}
This string corresponds, in the gauge theory on the branes, to a charge
in the fundamental representation of the gauge group and its energy
corresponds to the self energy of the charge.
This is one example of an energy in the field theory
which is related to a (static) string configuration in the geometric
description. We will encounter such situations again later.
Following \cite{Mald9711},\cite{Mald3002}, we identify the energy $E$
(in the field theory) corresponding to a static string configuration with
the (minimal) action of the string per unit time.
In the present situation, the action is
\[ S = \frac{1}{2\pi l_s^2} \int d\tau d\sg
\sqrt{\det_{\al\bt} G_{\mu\nu} \pt_\al X^\mu \pt_\bt X^\nu} \]
so, for a static configuration $X^0=\tau$, $\vec{X}=\vec{X}(\sg)$,
the corresponding energy is
\beql{E-F1-def}
E= \frac{1}{2\pi l_s^2} \int d\sg \sqrt{G_{00}G_{ij} {X'}^i {X'}^j} \hs.
\eeq
\internote{A static configuration exists since the metric is static.}
This is the length of the string in the {\em effective} metric
\[ dE^2=\frac{1}{4\pi^2 l_s^4} G_{00}ds^2 \]
and, for extremal D2 branes,
\beql{dE} dE^2 =
\left(\frac{\sin\bt}{2\pi N_6}\right)^2 \left[\frac{U^4}{32\pi^2
N_6 N_2} dx_\prl^2 +  \left(\frac{1}{4} dU^2 + U^2 d \tilde\Om_6^2
\right)\right] \hs. \eeq
\internote{For the non-extremal case, $dx_\prl^2$ and $d\tilde\Om_6^2$ are
multiplied by $f_h$.}
Observe that $N_6$ and $N_2$ can be ``absorbed'' in $dE$ and $x_\prl$,
so one can make the calculation with $N_2=N_6=1$ and then
recover $N_2$ and $N_6$ by
\[ E \goto N_6 E \hsc
x_\prl \goto \frac{x_\prl}{\sqrt{N_6 N_2}} \hs. \]

We return now to the present case -- a fundamental string with one end on the
D2 branes and the other extending to infinity -- where the corresponding energy
is the self-energy $E_F$ of a charge. To obtain a finite energy, we need an
IR cutoff in the geometry and a UV cutoff in the field theory, so this will
lead to a relation between the two.
An IR cutoff in the geometry means that the string extends to a finite point
$(U,\Om)$.%
\ftl{D2probe}{This can be realized by placing at this point an additional D2
brane, parallel to the others. The additional D2 brane is considered as a probe
and its influence on the geometry is neglected.}
\internote{In the gauge theory, this corresponds to a $U(N_2+1)$ gauge
symmetry, broken to $U(N_2)$ by an expectation value for the scalars $X_i$.}
By symmetry, for a minimal string, only $U$ and $\bt$ vary,
so the effective metric (\ref{dE}) leads to
\beql{EF} E_F = \int \frac{\sin\bt}{2\pi N_6}
\sqrt{\frac{1}{4} dU^2 + U^2 d\bt^2} \hs.
\eeq
We see that $E_F$ depends only on $U,\bt$ and $N_6$ and, moreover,
$N_6 E_F$ is independent of $N_6$. Therefore, considering dimensions,
we obtain for the self energy the following functional form
\beql{m-self-geo} E_F=\Ec_F\frac{U}{N_6} \hs. \eeq
\internote{For $m_Q>0$: $\Ec_F$ depends on $\bt, \th, \frac{U_Q}{U}$.}
$\Ec_F$ depends on $\bt$ and, in particular, it vanishes for
$\bt=0$. However, it is not clear how to relate this dependence to
the field theory on the D2 branes. The reason for this is that
$\bt\neq\frac{\pi}{2}$ corresponds to a string which, if extended
from the D2 branes, is not transverse to the D6 branes.
This is not a stable configuration%
\footnote{In particular, it preserves no supersymmetry.}
and it will ``decay'' to a string ending on the D6 branes.
In fact, in the metric (\ref{s-D2-n}), the D6 branes
totally coincide with the D2 branes (because of the $\sin\bt$ conformal
factor), so the string for which we calculated the action may very well
be one that ends on the D6 branes and {\em not } on the D2 branes.
The vanishing of $E_F$ for $\bt\goto0$ supports this identification.
If this is true,
we should not identify such a string with a charge in the gauge theory,
since this theory describes the dynamics of the D2 branes only.%
\internote{The situation changes when the D6 branes are compact.
\nl What should one say then?}
Because of the above complication, we will consider only $\bt=\frac{\pi}{2}$.
\internote{CAN $\Ec$ BE CALCULATED?}

To obtain a UV-IR relation, we need the cutoff dependence of the self-energy
of a charge in the field theory $E_F \propto \frac{1}{\dl x_\prl}$.
Comparing eqs. (\ref{UVIR-T}) and (\ref{m-self-geo}), we obtain the prediction
\beql{m-self-FT} E_F \sim \sqrt{\frac{N_2}{N_6}} \frac{1}{\dl x_\prl}
\hs. \eeq
Comparing to the corresponding result in the dynamics of D3 branes
$E_F\sim\frac{\sqrt{N} g_\YM}{\dl x_\prl}$ \cite{PP9809}, we see again%
\footnote{Recall that we observed this analogy in the previous subsection,
considering the quantum corrections to the geometric description.}
that $1/N_6$ here is analogous to $g_\YM^2$ there.

\newpar{Summary}

In this subsection, we considered the UV-IR relation from various points of
view and obtained the same results as obtained previously for D3 branes
\cite{SW9805},\cite{PP9809}: one can identify a relation
\[ \frac{1}{\dl x_\prl} \propto U_\max \]
between a UV cutoff $\dl x_\prl$ in the
gauge theory and an IR cutoff $U_\max$ in the geometry.
This cutoff defines two energy scales: one
\[ \frac{1}{\dl x_\prl} \sim \frac{U}{\sqrt{N_6 N_2}} \hs, \]
\internote{Combining $\frac{1}{\dl x_\prl}\propto U$ (required by dimensional
analysis) with any consideration that uses only the metric (\ref{s-D2})
and is not sensitive to a constant conformal factor, leads to
$\frac{1}{\dl x_\prl} \sim \frac{1}{R}\left(\frac{r_\max}{R}\right)^2$,
\nl where $R\sim(N_6 N_2)^{\frac{1}{6}} l_p$ comes from
$f_2=1+\left(\frac{R}{r}\right)^6$.}
relevant to thermodynamics and supergravity fields and another
\[ E_F \sim \sqrt{\frac{N_2}{N_6}} \frac{1}{\dl x_\prl}
\sim \frac{U}{N_6} \hs, \]
relevant to fundamental strings.

\subsection{The Potential between Fundamental Charges}

A charge in the fundamental representation of the gauge group -- a quark --
can be realized by an additional D2 brane, considered as a probe,
and by stretching a fundamental string between this probe and the $N_2$
D2 branes (see footnote \ref{D2probe}).
A configuration of a quark and an anti-quark can be
realized by two such D2 branes.
When, in the field theory, the quarks are very far from each other,
the energy is the sum of the self energies of the two quarks.
Geometrically, this energy corresponds to two fundamental strings
(with opposite orientation),
each connecting one of the additional D2 branes to the $N_2$ D2 branes.
When the distance between the fundamental strings is decreased,
it may be energetically favorable for the two strings to join to a single
string, connecting directly the two D2 branes. In the field theory, this would
correspond to an attractive potential between the charges \cite{RY3001}%
\cite{Mald3002}. The potential is given by eq. (\ref{E-F1-def}),
after subtracting the self energy (\ref{m-self-geo}) of the
quarks.
We are, therefore, led to consider a string stretched between
$(x_1, U,\Om_1)$ and $(x_2,U,\Om_2)$, where $\Om$ represents
the angular coordinates%
\footnote{Although everything said below is true for any $\bt$, one should
keep in mind that, as explained in subsection \ref{s-en}, we expect relevance
to the D2 brane dynamics only for $\bt=\frac{\pi}{2}$.}.
Scaling considerations, as in the derivation of eq. (\ref{m-self-geo}),
lead to the following functional form for the energy:
\[ E=\sqrt{\frac{N_2}{N_6}} \frac{1}{\Dl x}
   \Ec\left(\frac{U\Dl x}{\sqrt{N_2N_6}}\right) \hs, \]
where the function $\Ec$ depends also on the angles. As explained
above, we know that for $\Dl x \goto\infty$ (with a fixed cutoff
$U$) the energy $E$ remains finite, being the self energy of the
two quarks. Therefore, the expansion of $E$ for large $U \Dl x$
is:
\[ E = \frac{U}{N_6} \Ec_1 + \sqrt{\frac{N_2}{N_6}} \frac{1}{\Dl x}
   \left[ \Ec_0 + \Oc\left(\frac{1}{U\Dl x}\right)\right] \hs. \]
The first term is the self energy
(being independent of $\Dl x$ and proportional to $U$).
Subtracting it and removing the cutoff (taking $U$ to infinity)%
\ftl{f-UV}{Strictly speaking, the limit $U\goto\infty$ is allowed
only after taking $l_s\goto0$ and $g_\YM\goto\infty$ (see the end
of subsection \ref{s-phase-0}). Practically, what we want is to
suppress the $\Oc\left(\frac{1}{U\Dl x}\right)$ terms and these
will be suppressed also when $U$ is finite but $\Dl x$ is large
enough. In physical terms, the restriction on $U$ corresponds to
an effective UV cutoff in the field theory and this becomes
irrelevant at large distances.},
we obtain for the potential
\beql{V} V = \sqrt{\frac{N_2}{N_6}} \frac{1}{\Dl x} \Ec_0 (\Om_1,\Om_2) \hs.
\eeq
To find the function $\Ec_0$, one has to solve the geodesic equations of the
metric (\ref{dE}), which we do not do here. However, much can be deduced
already from the present form.
For negative $\Ec_0$, the potential describes a Coulomb-like (attractive) force
(as expected from conformal invariance),
with an effective charge which depends on the angles.
\internote{Recall that these coordinates correspond to internal degrees
of freedom in the gauge theory.}
When $\Ec_0>0$, apparently corresponding to a repulsive potential,
this actually means that the potential vanishes (classically), the
minimal string configuration being the disconnected one -- two strings
ending on the $N_2$ D2 branes.
\internote{In particular, this is what happens for
$\bt_1=\bt_2=0$, where $\Ec_0=0$.
\nl To see this, observe that the metric (\ref{dE}) degenerates so,
for finite $U$, $\bt_1=\bt_2=0$ imply that $(x_1, U,\Om_1)$ and
$(x_2,U,\Om_2)$ are the same point and, therefore, $\Ec$ vanishes.}
The vanishing of the potential means that the force between the quarks is
screened. From the point of view of the field theory, this is the expected
result, since the theory contains massless dynamical matter in the
fundamental representation of the gauge group -- the excitations of the
D2-D6 strings.
Screening typically leads to a non-vanishing,
exponentially-decreasing potential. In the present geometry such a
potential is ruled out by conformal invariance but this only means
that we are in the extreme IR, \ie, considering very large
distances, where this potential is negligible (see in this context
footnote \ref{f-UV}).
A vanishing potential was obtained also in \cite{GO5129}, and
there it was argued that an exponential correction arises as a
quantum effect, related to a singularity in the geometry. Here
such a correction can appear already classically, when considering
higher energy scales, corresponding to regions farther from the D6
branes.

\internote{In 11D, there is no topologically stable string that could
represent a confining flux tube.
\nl This, however, does not necessarily have implications for the stability
of a string in 10D. Consider, for example, a string in the presence of a
(distant) D6.}

\newsection{D2 Branes Localized Near D6 Branes}
\secl{s-D2-Q}

In this section we extend some of our analysis in the previous section
to D2 branes which are at non-vanishing distance from the D6 branes.
This corresponds to a non-vanishing mass $m_Q$ for the quarks $Q$.
The solution we obtain interpolates between the $m_Q=0$ solution --
far from the D2 branes -- and the $AdS_4 \times S^7$ geometry --
close to the D2 branes.
Since the distance from the brane represents the scale in the gauge theory,
this is a geometric realization of the renormalization-group flow%
\footnote{A geometric realization of an RG flow in field theory
was also considered in \cite{GPPZ},\cite{DZ10206}.}
from a conformal fixed point with massless quarks to a fixed point
without the quarks. We also find, for $N_6\gg1$ an intermediate
region, corresponding to energy scales where massive quarks are
relevant. We combine this information with the results of
subsection \ref{s-phase-0}, to a unified phase structure,
summarized in the introduction.
\internote{More motivation for this extension is given in the supplements} 

\subsection{The Geometry}

We consider (extremal) coinciding $M_2$ branes, positioned at
\[ r=r_Q \hsc \bt=\frac{\pi}{2} \hsc \th=0 \hsc \psi=0\mod2\pi \hs, \]
where $r_Q$ is the distance from the singularity.
In the covering space of the orbifold,
where $\psi$ ranges from 0 to $2\pi N_6$, we have $N_6$ images of $N_2$
M2 branes, located at the positions $\psi = 2\pi k$ for $k=1, \ldots, N_6$
(see Appendix \ref{sec-D6} for more details).
The corresponding metric is, therefore,
\beql{s-M2-Q}
ds_{11}^2=f_2^{-\frac{2}{3}}dx_\prl^2
+f_2^{\frac{1}{3}}(dr^2 + r^2 d\Om_7^2) \hs,
\eeq
where
\beql{f2-Q}
f_2 = 1 + 32\pi^2 N_2 l_p^6 \sum_{k=1}^{N_6} \frac{1}{r_k^6}
\eeq
and $r_k$ is the distance (in the flat covering space) to the $k$'th image.
In the Cartesian coordinates for the
covering space:
\[ z_1=r\sin\bt\cos\frac{\th}{2}e^{i\frac{\psi}{N_6}} \hsc
z_2=r\sin\bt\sin\frac{\th}{2}e^{i(\vph+\frac{\psi}{N_6})} \hs, \]
the images are at
\[ z_1 = r_Q e^{\frac{2\pi ik}{N_6}} \hsc z_2=0 \hs, \]
therefore,
\begin{eqnarray}\nonumber
r_k^2 & = & (r\cos\bt)^2 + |z_2|^2 + |z_1 - r_Q e^{\frac{2\pi ik}{N_6}}|^2 \\
\label{r_k} & = & r^2 +r_Q^2 - 2r r_Q \sin\bt \cos\frac{\th}{2}
            \cos\left(\frac{\psi-2\pi k}{N_6}\right) \hs.
\end{eqnarray}

\newpar{The Near Horizon Geometry}

Near the M2 brane horizon (when $f_2\gg1$),
the metric (\ref{s-M2-Q}) simplifies to
\beql{s-M2-Q-n}
ds_{11}^2 = l_p^2
\left[\frac{\wbar{U}^2}{(32\pi^2 N_6 N_2)^\frac{2}{3}} dx_\prl^2 +
(32\pi^2 N_6 N_2)^\frac{1}{3} \frac{U}{\wbar{U}}
\left(\frac{dU^2}{4U^2} + d\tilde\Om_7^2 \right)\right] \hs,
\eeq
where
\beql{Ubar-def} \frac{1}{\wbar{U}^3} = \frac{1}{N_6} \sum_k \frac{1}{U_k^3}
\eeq
and $U$ and $d\tilde\Om_7^2$ are as in (\ref{s-M2-n}),(\ref{U-def-M2}).
Correspondingly, the 10D geometry becomes
\beql{s-D2-Q-n}
ds_{10}^2 = l_s^2 \, \frac{\sin\bt}{N_6}
\left[\sqrt{\frac{U\wbar{U}^3}{32\pi^2 N_6 N_2}} dx_\prl^2
+ \sqrt{32\pi^2 N_6 N_2} \left(\frac{U}{\wbar{U}}\right)^\frac{3}{2}
\left(\frac{dU^2}{4U^2}+d \tilde\Om_6^2 \right)\right] \hs,
\eeq
\beql{phi-Q-n}
e^\phi = \left( \frac{32\pi^2 N_2}{{N_6}^5}\right)^\frac{1}{4}
(\sin\bt)^\frac{3}{2} \left(\frac{U}{\wbar{U}}\right)^{\frac{3}{4}} \hs,
\eeq
where $d\tilde\Om_6^2$ is as in (\ref{s-D2}).

\subsection{The Phase Structure}
\secl{s-phase-Q}

\newpar{The Renormalization Group Flow}

The transition between the high and low energy regimes in the field
theory occurs at energy scales which depend on the mass $m_Q$ of the
quarks and this mass is related to the D2-D6 distance $r_Q$.
Thus, before considering the RG flow, we first determine this
relation.
The quarks $Q$ in the gauge theory are related to the D2-D6 fundamental
strings, so the mass $m_Q$ of these quarks is related to the energy carried
by these strings.
This is again a situation where an energy in the gauge theory corresponds to
a string configuration, so we use eq. (\ref{E-F1-def}) to calculate it.
The string extends from the singularity, $U\sin\bt=0$, to the location
of the D2 branes (the horizon): $U=U_Q$, $\sin\bt=\cos\th=1$.
For a minimal string, only $U$ and $\bt$ vary, thus, eqs.
(\ref{E-F1-def}) and (\ref{s-D2-Q-n}) give
\[ m_Q = \int \frac{\sin\bt}{2\pi N_6}
\sqrt{\frac{1}{4} dU^2 + U^2 d\bt^2} \hs. \] This expression is
identical to that for the self energy considered in subsection
\ref{s-en} (eq. (\ref{EF})), so the answer is given by eq.
(\ref{m-self-geo}) with $U=U_Q$: \beql{mQ} m_Q = \Ec_F
\frac{U_Q}{N_6} \hs. \eeq

We turn now to the RG flow. At low energies, the quarks $Q$
decouple and the theory flows to the IR fixed point of pure $N=8$
SYM. In the geometric description, this should correspond to a
region close to the M2 branes, where the existence of the D6
branes is irrelevant and the geometry is $AdS_4\times S^7$. The
condition for this is that the sum in eq. (\ref{f2-Q}) is well
approximated by $1/r_*^6$, where $r_*$ is the distance to the
nearest D2 image, and this is the case iff $r_*\ll r_Q/N_6$ (since
$r_Q/N_6$ is the distance, in the covering space, between
neighboring images). We conclude that the regime of the above IR
fixed point is $U_* \ll m_Q/N_6$ (where we define, as usual,
$U_*=r_*^2/l_p^3$).

At the other extreme --  high energy scales --
the mass $m_Q$ is expected to be irrelevant.
In the geometric description (eq. (\ref{s-M2-Q-n})), $r_Q$ enters through
$\wbar{U}$.
For $r_*\gg r_Q/N_6$, the sum in (\ref{Ubar-def}) is well approximated by an
integral, leading to
\beql{Uxy} \left(\frac{U}{\wbar{U}}\right)^3=
\frac{(1+x^2)^2 + 2x^2 y^2}{[(1+x^2)^2 - 4x^2 y^2]^{\frac{5}{2}}} \hsc
x^2=\frac{U_Q}{U} \hsc y=\sin\bt \cos\frac{\th}{2}  \hsc
U_Q = \frac{r_Q^2}{l_p^3} \hs. \eeq
\internote{
$f_2 = 1 + \frac{32\pi^2 N_6 N_2 l_p^6}{r^6}
   \frac{(1+x^2)^2 + 2x^2 y^2}{[(1+x^2)^2 - 4x^2 y^2]^{\frac{5}{2}}}$,
\nl $x=\frac{r_Q}{r}$, $y=\frac{|z_1|}{r}=\sin\bt \cos\frac{\th}{2}$.}
Expanding for small $x$, we obtain
\beql{UUb} \left(\frac{U}{\wbar{U}}\right)^3 = 1 - 3x^2 (1-4y^2) + \Oc(x^4)
\eeq
\internote{$f_2 = 1 + 32\pi^2 N_6 N_2 \left(\frac{l_p}{r}\right)^6
[1 - 3x^2 (1-4y^2) + \Oc(x^4)]$}
so, the mass is irrelevant when $x\ll0$, which means
$N_6 m_Q\ll U \approx U_*$.
In this region, the geometry is as for $m_Q=0$, which was considered in the
previous section.
Comparing the two extremes, one observes that for large $N_6$, there is an
intermediate region $m_Q/N_6 \ll U_* \ll N_6 m_Q$, in which the M2 branes are
effectively smeared along the circle parametrized by $\psi$.
\internote{Such an intermediate region appears also for D2's without D6's
\cite{IMSY9802}: $g_\YM^2 N_2^{\frac{1}{5}} \ll U \ll g_\YM^2$.}

In the field theory, the mass is expected to be negligible whenever it is
small compared to the energy considered (\ie, one does not expect factors of
$N_2$ or $N_6$ in this relation). With this in mind, the geometry suggests
that when we consider quarks, the energy scale in the field theory,
corresponding to the location $U_*$, is
\beql{EU} E\sim \frac{U_*}{N_6} \hs. \eeq
\internote{This is established only for $f_2,f_6\gg1$!}
Note that for large $U$, when the geometry is approximated by that of massless
quarks, this identification coincides with that suggested in the previous
section (see eq. (\ref{m-self-geo})).

Turning now to the range of the $AdS_4\times S^7$ geometry, the
identification (\ref{EU}) implies that the condition for the
decoupling of the quarks is
\[ E\ll \frac{m_Q}{N_6^2} \hs. \]
The $N_6$ dependence is reasonable: the influence of the
quarks (as intermediate states) grows with $N_6$, therefore, the energy at
which they decouple gets smaller.
\internote{IS THERE A MORE QUANTITATIVE INFORMATION?}

\newpar{Reduction to 10 Dimensions}

Next we check the transition between 11 dimensional and 10 dimensional
geometries. As before, this is governed by the
dilaton $\phi$, which for the present case is given in eq. (\ref{phi-Q-n}):
\[ \frac{R_\psi}{l_p} = e^{2\phi/3}
= \left(\frac{32\pi^2N_2}{N_6^5}\right)^\frac{1}{6} \sin\bt
\sqrt{\frac{U}{\wbar{U}}} \hs. \]
For $r_Q/N_6 \ll r_*$, \ie, when it is valid to approximate the sum in eq.
(\ref{f2-Q}) by an integral (see eq. (\ref{Uxy})), the geometry is independent
of $\psi$. This means that
\[ \cos\left(\frac{\psi-2\pi k_*}{N_6}\right)\approx1 \hs, \]
so eq. (\ref{r_k}) gives
\[ \left(\frac{r_*}{r_Q}\right)^2 \approx
\frac{\dl x^2 + 2(1 + \dl x) \dl y}{(1 + \dl x)^2} \hsc
\dl x = x-1 \hsc \dl y = 1-y \hs, \]
with $x,y$ defined in (\ref{Uxy}). This implies
\[ r_* \ll r_Q \hs\hs \Longleftrightarrow \hs\hs \dl x^2, \dl y \ll 1 \hs. \]
Combining this with eq. (\ref{Uxy}) we obtain,
for $r_Q/N_6 \ll r_* \ll r_Q$,
\internote{Since $r_* \ll r_Q$ implies $U \approx U_Q$, $\sin\bt \approx 1$.}
\[ \left(\frac{U}{\wbar{U}}\right)^3 \approx 6 \left(\frac{r_Q}{2r_*}\right)^5
= \frac{6}{32} \left(\frac{U_Q}{U_*}\right)^{\frac{5}{2}} \hs, \]
\internote{$f_2-1 \approx \frac{32\pi^2 N_6 N_2 l_p^6}{r_Q^6}
\cdot 6 \left(\frac{r_Q}{2r_*}\right)^5
= 6\pi^2 N_6 N_2 \frac{l_p^6}{r_Q r_*^5}$.}
which leads to
\beql{Rpsi-int} \frac{R_\psi}{l_p} \approx \left[6\pi^2 N_2
\left(\frac{m_Q}{N_6 U_*}\right)^{\frac{5}{2}}\right]^\frac{1}{6} \hs. \eeq
Thus, as $U_*$ is decreased, $R_\psi$ increases. As discussed above, we
consider $N_2\gg1$, therefore,
by the time $U_*$ reaches $m_Q/N_6$
(where the M2 branes are not smeared anymore),
$R_\psi/l_p$ is large and the geometry is 11 dimensional.
\internote{WHAT IS THE CURVATURE FOR $m_Q>0$? (for 10 and 11 dimensions).}

\newpar{Summary}

The phase structure that emerges from the discussion in this
subsection and subsection \ref{s-phase-0}, is described in figure
\ref{f-phs} (in the introduction).
It is assumed that $N_2\gg1$, otherwise, as we saw, there is no situation
with small curvature.
The coordinates are $N_6$ and $E=U_*/N_6$ and, as discussed above, we
identify $E$ with the energy scale in the field theory%
\footnote{Recall that this is the identification relevant to processes
involving fundamental strings. Note also that this coincides with the
identification used in \cite{IMSY9802} -- the action (per unit of time) of
a static string.}.
\internote{Supplements:
\nlb Compare to \cite{IMSY9802}:
\nl Taking $\sin\bt=1$ and identifying $r_2^2$ with $\rho^2$ of eq.
(\ref{rho-def}), one obtains:
\nl $\frac{U}{N_6} = \frac{r_2^2}{N_6 l_p^3} =
\frac{r_2^2}{N_6 g_s l_s^3} = \frac{2r_6}{l_s^2}$
\nlb The alternative $E=\frac{U_*}{\sqrt{N_6 N_2}}$:
\nl The only change in the
diagram is in the line $g_\eff=1$ (relative to the other lines).
\nl WHAT IS THE CORRECT CHOICE?}

At scales below $m_Q$ we
arrive at the conformal theory of the pure $N=8$ SYM, corresponding to the
geometry $AdS_4\times S^7$. Going up in energy, we pass through a region
where the geometry is still 11 dimensional but the M2 branes are effectively
smeared. This region exist only for large $N_6$ and corresponds to a regime
in the field theory where the massive quarks have significant influence.
At scales above $m_Q$, the mass is negligible and there are the three phases,
described in subsection \ref{s-phase-0}, depending on
the relation between $N_2$ and $N_6$.
Going further up in energy, one gets to the scale where the effective
gauge coupling $g_\eff=\frac{N_2 g_\YM^2}{E}$ becomes small and there
is a weakly coupled field theoretical description. Before this happens,
$f_6$ stops being large and the geometric description includes also the
geometry of isolated D2 branes.
Finally, at a scale determined by the string scale,
there is a further transition to an energy regime where stringy
dynamics is important.
\internote{Additional Issues:
\nlb The Potential between fundamental charges:
\nl eq. (\ref{V}) generalizes to
\nl $V = \sqrt{\frac{N_2}{N_6}} \frac{1}{\Dl x}
\Ec_0\left(\sqrt{\frac{N_6}{N_2}} m_Q \Dl x, \Om_1,\Om_2\right)$.
\nl CAN WE SAY SOMETHING ABOUT THE $m_Q$ DEPENDENCE?
\nl For $m_Q>0$, assuming analyticity at $m_Q=0$,
finiteness for $\Dl x\goto\infty$ seems to allow only
a linear term in $\frac{N_6}{N_2} m_Q \Dl x$.
However, there seems to be no reason for analyticity at this singular point.}

\section{Relations to Other Systems}

\secl{s-dual}

The D2-D6 system analyzed in this work is related, by
compactification and U-duality, to several other systems. There
are qualitative differences between some of these systems and in
this section we check how these differences arise in the geometric
and field-theoretic descriptions. We find satisfactory agreement
and this should be considered as a (successful) consistency check
of the proposed AdS-CFT duality. It is important to keep in mind
that duality is, by definition, a relation between two
descriptions of {\em the same} system, so U-duality cannot account
for the differences mentioned above. The essential change is,
therefore, compactification (in one of the dual descriptions).

\newpar{Compactification and the RG Flow}

We start by compactifying one of the directions of the D2 brane.
The corresponding field theory is on $\RR^2 \times S^1$ and it
flows from a 3 dimensional theory in the UV to a 2 dimensional
theory in the IR. In the geometric description, taking $x_2$ to
be periodic with period $2\pi R_2$, the physical compactification
radius is $\sqrt{G_{22}}R_2$. Since $G_{22} \sim U^2$ (see eq.
(\ref{s-D2-n})) and $U$ is proportional to the energy scale in the
field theory, we indeed obtain a 3 dimensional world-volume at
high energies and a 2 dimensional one at low energies. Note that
we only used the fact that the metric components in directions
parallel to the world-volume increase with $U$, which is true in
all $AdS$ geometries, so this correct realization of the
RG flow will exist in any AdS-CFT relation%
\footnote{Moreover, it also appears to exist in non-conformal
D$p$-branes analyzed in \cite{IMSY9802}, although there the
geometric identification of the energy scale is more problematic,
as discussed in \cite{PP9809}.}.

\newpar{The D1-D5 System and Localization}

Performing T-duality in the (compactified) $x_2$ direction, one obtains D1
branes {\em localized} on D5 branes. However, the geometry obtained does not
have the SO(2,2) symmetry required to identify it with a conformal theory
on the D1 branes. Instead, such a geometry is obtained \cite{Mald9711},
when the D1 branes are {\em smeared} over the D5 branes. In the D2-D6
system, the situation is reversed and, as we saw in section \ref{s-D2-0},
a geometry with $SO(2,3)$ symmetry is obtained only when the D2 branes are
fully localized on the D6 branes.
This difference has a clear counterpart in field theory.
For localized branes, there is a geometric moduli space of their possible
positions, which corresponds, in the field theory,
to the (classical) moduli space of vacua.
In three dimensions (and more), the vacuum of a field theory
corresponds to specific values of the moduli and, therefore, the corresponding
geometry should be that of localized branes%
\footnote{Continuously smeared branes were considered in \cite{KLT11120},
but this was meant as an approximation to a specific distribution of branes,
which means, a specific point in the moduli space of vacua.}.
In two dimensions (and less), there is tunneling between
continuously-connected classical ground states \cite{MW66}\cite{coleman},
leading to a single quantum ground state and this is
reflected in the geometry by smearing. Note that the smearing is
in directions parallel to the D5 branes, which correspond to the
Higgs branch of the moduli space and there is no smearing in the
other directions, corresponding to the Coulomb branch. This is in
agreement with other evidence \cite{Witten9707} that the
corresponding world-volume theory is in the Higgs branch.

\newpar{The D3-D7 System and a Limit on the Number of Flavors}

Next, we consider a compactification of a direction transverse to
the D6 branes. This leaves two non-compact transverse directions
and in such a situation, the number $N_6$ of D6 branes is limited.
Such a bound exists also in the field-theoretical description:
T-dualizing along the compact direction, one obtains D3 branes
localized on D7 branes, with one of the D3 brane world-volume
directions compactified on a dual circle. The corresponding field
theory is a 4 dimensional gauge theory on $\RR^3 \times S^1$. The
number $N_6$ of D7 branes is the number of flavors in the gauge
theory and requiring asymptotic freedom, one obtains a bound on
$N_6$. To remove the bound, the radius of the dual circle must
strictly vanish (since asymptotic freedom is a property of the
small distance behavior) and this corresponds to D6 branes with 3
non-compact transverse directions. Therefore, we see that both
approaches lead to the same kind of limit on $N_6$.
To obtain a detailed agreement, further considerations are needed%
\footnote{Note that a straight-forward application of the
asymptotic freedom requirement in the present case would role out
any $N_6>0$.} and this will not be done here.
\internote{ \nlb Actually, this theory is not asymptotically free
for any $N_6$ (since it is finite already for $N_6=0$). \nlb
Compactifying the remaining transverse dimensions, $N_6$ is
limited to 32, (with 4 O7 planes) and the theory on the D3 branes
can be seen as a finite $Usp(2N_2)^4$ gauge theory, with the gauge
theory broken by vev's. \nlb In non-compact transverse dimensions,
$N_6$ is still restricted, since each D7 brane leads to a deficit
angle $2\pi/24$ in the transverse space. The resulting bound is
$N_6\le24$.}

\newpar{A Compactified Orbifold and Mirror Symmetry}

Finally, we consider the compactification of D6 directions which are transverse
to the D2 brane. To be specific, let $x^6$ be periodic, with period
$2\pi R_6$. When $R_6 \ll l_p$, one can return to 11 dimensions and identify
$x^6$ as the small 11th dimension. This leads to a configuration of $N_2$
D2 branes on a $\ZZ_{N_6}$ orbifold.
The field theory on the D2 branes is
$\prod_{\al=1}^{N_6} U(N_2)_\al$ gauge theory with matter in the
$(N_2,\wbar{N_2})$ representation of $U(N_2)_\al \otimes U(N_2)_{\al+1}$ for
each $\al$ (with cyclic identification $\al\sim\al+N_6$)
and an additional fundamental for one of the $U(N_\al)$ factors
\cite{DoMo9603}.
Thus, for large $N_6$ and small $R_6$ there are two field theoretical
descriptions of the same system and indeed,
it is known that these two field theories are
related by mirror symmetry \cite{InSe9607}\cite{mirror}.

\newpar{A Compact D6 brane}

When all the D6 directions (which are transverse to the D2 branes)
are compactified, the dynamics of the D6 branes cannot be ignored
and the corresponding field theory is a $U(N_2) \otimes U(N_6)$
gauge theory.  The $U(N_6)$ gauge coupling is inversely
proportional to $V^\frac{1}{8}$ (where $V$ is the compactification
radius) so, for infinite $V$, it vanishes and the $U(N_6)$
symmetry is a global symmetry. This seems as a smooth limit, but
actually, there is a conceptual difference between global and
gauge symmetries, the former being a real symmetry, while the
later is a redundancy in the description. In the string theory
context, this sharp distinction is reflected by the fact that
global symmetries in the brane's world-volume theory correspond to
gauge symmetries (and gauge fields) in the bulk, while gauge
symmetries on the brane are not seen in the bulk. We, therefore,
return to our model and check how this distinction is seen.

To obtain the geometry, one starts, as before, in the (11 dimensional)
covering space, and considers an array of M2 branes located at the points
of the lattice $\Gm$ defined by the compactification (\ie, $\RR^4/\Gm$ is the
compactification torus). The corresponding metric is
\beql{s-M2-C}
ds_{11}^2=f_2^{-\frac{2}{3}}dx_\prl^2 + f_2^{\frac{1}{3}} d\vec{r}^2 \hsc
f_2 = 1 + 32\pi^2 N_6 N_2 l_p^6 \sum_{k\in\Gm} \frac{1}{r_k^6} \hs,
\eeq
where $d\vec{r}^2$ is the (flat) metric of the transverse covering space,
and $r_k$ is the distance to the lattice point $k$.
Following the usual procedure, we obtain the 10 dimensional near-horizon
geometry%
\footnote{We use here, for the transverse space, the same
spherical coordinates as in the previous section. To analyze
distances larger than the compactification lengths, one should use
other coordinates, but this will not be done here explicitly.}
\beql{s-D2-C-n} ds_{10}^2 = \frac{l_s^2}{N_6}
\sqrt\frac{U}{\wbar{U}} \sin\bt
\left[\frac{\wbar{U}^2}{\sqrt{32\pi^2 N_6 N_2}} dx_\prl^2 +
\sqrt{32\pi^2 N_6 N_2}\frac{U}{\wbar{U}}
\left(\frac{dU^2}{4U^2}+d\tilde\Om_6^2\right)\right] \hs, \eeq
\beql{phi-C-n} e^\phi = \left[\frac{32\pi^2 N_2}{N_6^5}
\left(\frac{U}{\wbar{U}}\right)^3 \sin^6\bt\right]^\frac{1}{4}
\hs, \eeq where \beql{U-C-def} U=\frac{r^2}{l_p^3} \hsc
\frac{1}{\wbar{U}^3} = \sum_k \frac{1}{U_k^3} \hs. \eeq An $AdS$
geometry is obtained only for $r$ small compared to the
compactification length's, (when $\wbar{U}\approx U$), which
corresponds to low energy scales in the world-volume theory. This
agrees with field-theoretical expectations, since at larger
scales, the 6-dimensional physics is no longer negligible.
\internote{For large $U$ one obtains the solution with smeared D2 branes.
See the supplements for details.}
The AdS geometry (for small $U$) is locally the same as for
the non-compact D6 branes. In particular, there is a singularity at
$\bt=0$ and, as always, one expects that it corresponds to additional
massless degrees of freedom living there. For non-compact D6 branes we
identified them as the open strings ending on these branes, and the
implied $U(N_6)$ gauge symmetry in the bulk%
\footnote{In this context, ``bulk'' means ``off the D2 branes'';
of course, the $U(N_6)$ gauge fields are on the D6 branes and not in the
10 dimensional bulk.}
agreed with the global symmetry in the world-volume theory.
However, in the present situation, the $U(N_6)$ symmetry is {\em a
gauge symmetry in the world-volume theory}, so the singularity
must have a different origin. To find what it is, one observes
that the length of the compact directions vanishes as $\bt$ is
decreased to 0 (because of the $\sin\bt$ factor in the metric), so
the massless degrees of freedom at the singularity are {\em closed
strings} winding around these directions. When only part of the 4
extra dimensions of the D6 branes are compactified, the
singularity is due to both closed and open strings. In all these
cases we have agreement between the global symmetry in the
world-volume theory and the gauge symmetry in the bulk.

\vspace{1cm}
\noindent{\bf Acknowledgment:}

The initial stages of
this research were performed in collaboration with David Kutasov
and it is a pleasure to thank him for numerous suggestions and
comments.
We also thank Ofer Aharony, Finn Larsen and Emil
Martinec for helpful discussions.
The work of O.P. was supported by DOE grant DE-FG02-90ER-40560 and
NSF grant PHY 91-23780.


\appendix
\renewcommand{\newsection}[1]{
 \vspace{10mm} \pagebreak[3]
 \refstepcounter{section}
 \setcounter{equation}{0}
 \message{(Appendix \thesection. #1)}
 \addcontentsline{toc}{section}{
  App. \protect\numberline{\Alph{section}}{\hs\hs\boldmath #1}}
 \begin{flushleft}
  {\large\bf\boldmath Appendix \thesection. \hspace{5mm} #1}
 \end{flushleft}
 \nopagebreak}


\newsection{The Near-Horizon Geometry of  Coinciding D6 Branes}
\secl{sec-D6}

D6 branes are identified in M theory with KK monopoles \cite{Town}.
The corresponding transverse metric -- the Taub-NUT metric --
for $N_6$ coinciding D6 branes is
\beql{g-D6}
ds_{\rm TN}^2=f_6[dr_6^2+r_6^2d\Om_2^2]
+f_6^{-1}R_\#^2[d\psi+\half N_6(1-\cos\th)d\vph]^2 \hs,
\eeq
where
\beql{f6-def} f_6=1+\frac{N_6R_\#}{2r_6} \hsc R_\#>0 \hs, \eeq
$d\Om_2^2$ is the metric on the unit 2-sphere
\[ d\Om_2^2=d\th^2+\sin^2\th d\vph^2 \hsc \]
($0\le\th\le\pi$; $\vph$ is periodic with period $2\pi$)
and the angle $\psi$ parametrizes the compact Taub-NUT direction,
normalized to have a periodicity of $2\pi$.

For {\em large} $r_6$ (far from the D6 branes; where $f_6\approx1$),
the 11D metric
approaches that of flat $\RR_\prl^{6+1}\times\RR_\perp^3\times S^1$,
$R_\#$ being the asymptotic radius of the compact dimension.
M theory on this space is identified with type IIA string theory
\cite{Witten9503}, and the string scale $l_s$ and the
string coupling $g_s$ are given in terms of the asymptotic radius
 $R_\#$ and the 11-dimensional Planck length $l_p$ by
\beql{lgs}
l_s^2=\frac{l_p^3}{R_\#} \hsc g_s^2=\left(\frac{R_\#}{l_p}\right)^3 \hs.
\eeq

For {\em small} $r_6$ (near the D6 branes; where $f_6\gg1$), the
transverse space (\ref{g-D6}) becomes an ALE-space with
$A_{N_6-1}$ singularity \cite{GiHa}. The corresponding metric
reads \beql{g-orb} ds_{\rm ALE}^2 = d\rho^2+\rho^2d\tilde\Om_3^2
\hs, \eeq where \beql{rho-def} \rho^2=2N_6R_\#r_6, \eeq and
\[ d\tilde\Om_3^2:=\frac{1}{4}d\Om_2^2
+\left[\frac{1}{N_6}d\psi+\frac{1}{2}(1-\cos\th)d\vph\right]^2 \hs. \]
This metric describes an orbifold $\RR^4/\ZZ_{N_6}$.
To demonstrate this, one changes coordinates
\beql{orb-coo} \th=2\hat\th \hsc \vph=\hat\vph_2-\hat\vph_1 \hsc
   \psi=N_6\hat{\vph}_1 \hs, \eeq
obtaining
\beql{s-orb-ang} d\Om_3^2:=d\hat\th+\cos^2\hat\th d\hat\vph_1^2
            +\sin^2\hat\th d\hat\vph_2^2 \hs. \eeq
Extending the periodicity of $\psi$ to $2\pi N_6$, one obtains that
$0\le\hat\th\le\frac{\pi}{2}$ and $\hat\vph_i$ are periodic with
period $2\pi$. So, defining
\beql{z-def} z_1=\rho\cos\hat\th e^{i\hat\vph_1} \hsc
   z_2=\rho\sin\hat\th e^{i\hat\vph_2} \hs, \eeq
the metric (\ref{g-orb}) becomes that of flat $\CC^2$:
\[ ds^2=|dz_1|^2+|dz_2|^2 \hs. \]
Since the range of $\psi$ is $2\pi$ (and not $2\pi N_6$) $\CC^2$ covers our
space $N_6$ times, with the following identifications:
\beql{orb-trans}
(z_1,z_2) \sim (\al z_1,\al z_2) \hs \forall\al^{N_6}=1 \hs. \eeq

\newsection{NS5 Branes Localized On D6 Branes}
\secl{sec-NS5}

The NS5 branes decouple from the stringy bulk
degrees of freedom when the string coupling $g_s$ vanishes
\cite{Sei9705}.
\internote{Details:
\nl The condition is $1\ll f_5-1=\frac{N_5 l_s^2}{r^2}$
\nlb type IIB:
$E_B\ll\frac{\sqrt{N_5}}{g_B l_s}=
\frac{\sqrt{N_5\tilde{g}_B}}{\tilde{l}_s}$
\nl where $E_B=U_B=\frac{r}{g_B l_s^2}=\frac{r}{\tilde{l}_s^2}$
(the mass of a D1/F1).
\nlb type IIA:
$E_A\ll\frac{N_5^{1/4}}{\sqrt{g_A}l_s}$
\nl where $E_A^2=\frac{U_A}{l_s}=\frac{r}{g_A l_s^3}$ (the tension of a D2).}

The resulting world-volume theory is expected to be a non-local 6 dimensional
theory without gravity \cite{Sei9705}.
It has stringy excitations (which can be thought of as boundaries of D2
branes ending on the NS5 branes) and is called ``little string theory''.
The tension of these strings is proportional to $1/l_s^2$.
In the absence of D6 branes, the theory has (2,0)
supersymmetry. It was analyzed, in the context of the AdS-CFT correspondence,
in \cite{IMSY9802}\cite{ABKS9808}.
The D6 branes break the supersymmetry to (1,0) and add
3 dimensional degrees of freedom -- the boundaries of D4 branes extending
between the NS5 branes and the D6 branes.
\internote{
\nl In M theory this is a deformation of the M5 branes induced by the
orbifold.
\nl An analog of $m_Q$: the worldvolume area of a minimal D4 (it is finite!).}

In this appendix we consider briefly the world-volume theory of
NS5 branes localized on D6 branes, as described above. One can
repeat most of the analysis of the D2-D6 system, performed in the
main text, also for the D5-D6 system and very similar results are
obtained. In particular, also here the geometry is that of a
warped product. So far, there is not much that is known about the
world-volume theory of decoupled NS5 branes, therefore, the
information provided by the present approach, as sketched below,
should be useful in the further study of this system.

\newpar{The Geometry}

The geometry near the horizon of D6 branes, in the presence of NS5 branes
was determined in \cite{ITY3103}. To reproduce it,
we follow the same steps as in section \ref{s-D2-0}.
For simplicity, we consider {\em extremal} NS5 branes {\em on} the D6
branes.
A NS5 brane corresponds to an M5 brane in 11 dimensions.
The metric corresponding to $N_6N_5$ coinciding
images of M5 branes in the (flat) covering space of the orbifold is
\beql{s-M5}
ds_{11}^2=f_5^{-\frac{1}{3}} dx_\prl^2
+ f_5^{\frac{2}{3}} (dr^2+r^2d\Om_4^2) \hs,
\eeq
\beql{f5-def} f_5=1+\frac{\pi N_6 N_5 l_p^3}{r^3} \hsc
d\Om_4^2=d\bt^2+\sin^2\bt d\Om_{3\perp}^2 \hs, \eeq
\internote{$A^{(6)} \equiv (A^{(3)})^* =
\frac{f_5-1}{f_5} dx^0 \wedge \ldots \wedge dx^5$.}
where $d\Om_4^2$ is the metric on the unit 4-sphere,
parametrized using the metric $d\Om_{3\perp}^2$ of a unit 3-sphere
and an additional angle $0\le\bt\le\pi$.
The orbifold metric is obtained, as in section \ref{s-D2-0}, by replacing
$d\Om_{3\perp}^2$ by the metric $d\tilde\Om_3^2$ on $S^3_\perp/\ZZ_{N_6}$
(see eq. (\ref{s-s3})).
The reduction to 10D gives
\beql{s-NS5} ds_{10}^2=e^{2(\phi-\phi_\infty)/3}
\{f_5^{-\frac{1}{3}} dx_\prl^2
+ f_5^{\frac{2}{3}}[dr^2+r^2 (d\bt^2+\frac{1}{4}\sin^2\bt d\Om_2^2)]\}
\eeq
and
\beql{phi-NS5}
e^\phi=g_s f_5^{\frac{1}{2}}
\left(\frac{r \sin\bt}{N_6 R_\#}\right)^{\frac{3}{2}} \hs.
\eeq
\internote{$e^\phi = g_s f_5^{\frac{1}{2}} f_6^{-\frac{3}{4}}$.}

Near the M5 brane horizon (when $f_5\gg1$), the metric (\ref{s-M5})
simplifies to
\beql{s-M5-n}
ds_{11}^2 = l_p^2 \left[\frac{U^2}{(\pi N_6 N_5)^{\frac{1}{3}}} dx_\prl^2
+ (\pi N_6 N_5)^{\frac{2}{3}}
\left(4 \frac{dU^2}{U^2} + d\tilde\Om_4^2\right)\right] \hsc
U^2=\frac{r}{l_p^3} \eeq
where $d\tilde\Om_4^2$ is the metric on $S^4/\ZZ_{N_6}$, as described above.
This is the space $AdS_7\times S^4/\ZZ_{N_6}$.
The 10D geometry becomes
\beql{s-NS5-n}
ds_{10}^2 = \frac{l_s^2}{N_6} \sin\bt
\left[U^2 dx_\prl^2 + \pi N_6 N_5
\left(4\frac{dU^2}{U^2}+d\bt^2+\frac{1}{4}\sin^2\bt d\Om_2^2\right)\right] \hs,
\eeq
\beql{phi-NS5-n}
e^\phi = \sqrt{\pi \frac{N_5}{N_6^2} \sin^3\bt} \hs.
\eeq
This is the geometry of an $AdS_7$ {\em fibered} over a compact manifold $X_3$
(parametrized by $\bt$ and the coordinates of $d\Om_2^2$),
where the radius of the $AdS_7$ space depends on orientation (the angle $\bt$)
\beql{R-AdS-NS5}
R_{\rm AdS}^{(10)} = 2 l_s \sqrt{\pi \frac{N_5}{N_6} \sin\bt} \hs.
\eeq

\newpar{The Phase Structure}

The dependence on the number of branes is qualitatively the same as in the
D2-D6 system: to obtain a classical geometric description (\ie, with small
curvature), $N_5$ must be large.
When this is satisfied, there are three phases, depending on
the relation between $N_5$ and $N_6$.
For $\sqrt{N_5} \ll N_6\ll N_5$ there is a geometric
description in 10 dimensions and for smaller $N_6$ there is such a description
in 11 dimensions. For larger $N_6$ there is no geometric description,
possibly indicating a simplification in some non-geometric
description. It would be interesting to consider this limit in other
approaches to this system.

Considering the dependence on energy, one first observes that, as in any
conformally-invariant geometry, the energy scale is proportional to $U$.
\internote{One can also restrict the $N_5$ and $N_6$ dependence.}
The restriction to the near-horizon range of the NS5 branes implies
\[ U^6 \ll \frac{N_6 N_5}{g_s^2 l_s^6} \hs, \]
which is satisfied in the decoupling limit $g_s\goto0$, so also in this case
the near horizon condition has the same meaning as in other systems.
The restriction to the near-horizon range of the D6 branes implies
\[ U^2\ll\frac{N_6}{l_s^2\sin\bt} \]
which is, as in the D2-D6 system, an independent condition
(\ie\ non-trivial for $g_s\goto0$). Its meaning is also the same: it
determines the domain of the IR fixed point which is, as expected, bounded by
the (little) string scale.
Finally, one can take the NS5 branes off the D6 branes and realize an RG
flow between two different fixed points, as was done for the D2-D6 system.

\ifinter\newpage\beginsup

\input{d2sup}

\endsup\newpage\fi


\end{document}